\documentclass{aa}  

\usepackage{graphicx}
\usepackage{lscape}
\usepackage{longtable}
\usepackage{natbib}
\usepackage{color}
\usepackage{array} 
\usepackage{array} 
\usepackage{tikz,array}
\usetikzlibrary{calc}

\bibpunct{(}{)}{;}{a}{}{,} 

\usepackage{txfonts}
\newcommand{\troy}{{\fontfamily{pzc}\fontsize{10pt}{13pt}\fontseries{b}\selectfont TROY}}
\newcommand{\troysection}{{\fontfamily{pzc}\fontsize{11pt}{13pt}\fontseries{b}\selectfont TROY}}

\newcommand\gO{{\cal O} }

\begin{document}

   \title{The {\fontfamily{pzc}\fontsize{19pt}{13pt}\fontseries{b}\selectfont TROY} project: Searching for co-orbital bodies to known planets}
   \subtitle{I. Project goals and first results from archival radial velocity}

   \author{
   J. Lillo-Box\inst{\ref{eso}}, 
   D. Barrado\inst{\ref{cab}}, 
   P. Figueira\inst{\ref{iace}}, 
   A. Leleu\inst{\ref{cheops}}\fnmsep\inst{\ref{obsparis}},
   N. C. Santos\inst{\ref{iace}}\fnmsep\inst{\ref{depfisporto}}, 
   A.C.M. Correia\inst{\ref{obsparis}}\fnmsep\inst{\ref{aveiro}},
   P. Robutel\inst{\ref{obsparis}},
   J. P. Faria\inst{\ref{iace}}\fnmsep\inst{\ref{depfisporto}}
          }

\institute{
European Southern Observatory, Alonso de Cordova 3107, Vitacura Casilla 19001, Santiago 19, Chile \label{eso}\\
              \email{jlillobox@eso.org}
\and Depto. de Astrof\'isica, Centro de Astrobiolog\'ia (CSIC-INTA), ESAC campus 28692 Villanueva de la Ca\~nada (Madrid), Spain\label{cab} 
\and Instituto de Astrof\' isica e Ci\^encias do Espa\c{c}o, Universidade do Porto, CAUP, Rua das Estrelas, PT4150-762 Porto, Portugal \label{iace} 
\and Departamento de F\'{i}sica e Astronomia, Faculdade de Ci\^{e}ncias, Universidade do Porto, Portugal   \label{depfisporto} 
\and CHEOPS fellow, Physikalisches Institut, Universitaet Bern, CH-3012 Bern \label{cheops}
\and IMCCE, Observatoire de Paris - PSL Research University, UPMC Univ. Paris 06, Univ. Lille 1, CNRS, 77 Avenue Denfert-Rochereau, 75014 Paris, France \label{obsparis} 
\and CIDMA, Departamento de F\'isica, Universidade de Aveiro, Campus de Santiago, 3810-193 Aveiro, Portugal \label{aveiro}
            }
            
  \titlerunning{The {\fontfamily{pzc}\fontsize{9pt}{13pt}\fontseries{b}\selectfont TROY} Project: I. Description, goals, and implications}
\authorrunning{Lillo-Box et al.}
   \date{\today}

\abstract
{ The detection of Earth-like planets, exocomets or Kuiper belts show that the different components found in the solar system should also be present in other planetary systems. Trojans are one of these components and can be considered fossils of the first stages in the life of planetary systems. Their detection in extrasolar systems would open a new scientific window to investigate formation and migration processes.}
{In this context, the main goal of the {\fontfamily{pzc}\fontsize{9pt}{13pt}\fontseries{b}\selectfont TROY}\ project is to detect  exotrojans for the first time and to measure their occurrence rate ($\eta$-Trojan). In this first paper, we describe the goals and methodology of the project. Additionally, we used archival radial velocity data of 46 planetary systems to place upper limits on the mass of possible trojans and investigate the presence of co-orbital planets down to several tens of Earth masses.}
{We used archival radial velocity data of 46 close-in ($P<5$ days) transiting planets (without detected companions) with information from high-precision radial velocity instruments. We took advantage of the time of mid-transit and secondary eclipses (when available) to constrain the possible presence of additional objects co-orbiting the star along with the planet. This, together with a good phase coverage, breaks the degeneracy between a trojan planet signature and signals coming from additional planets or underestimated eccentricity.}
{We identify nine systems for which the archival data provide $>1\sigma$ evidence for a mass imbalance between L$_4$ and L$_5$. Two of these systems provide $>2\sigma$ {detection, but no significant detection is found among our sample}. We also report upper limits to the masses at L$_4$/L$_5$ in all studied systems and discuss the results in the context of previous findings.}
 {}

   \maketitle
   
%
\section{Introduction}

After millennia of wondering, we now know that extrasolar planets abound \citep{pepe14,mayor14,batalha14, lissauer14b}. We have also proven several instances of exocomets  ($\beta$ Pic; \citealt{kiefer14}) and since 1984, with the Infrared Astronomical Satellite \citep[IRAS,][]{neugebauer1984}, we are also aware of Kuiper belt structures around other stars \citep[see][and references therein]{moro-martin08}. These discoveries imply that the non-planetary components of our solar system are not an exception but instead the rule, as these components are also present in extrasolar systems as a result of the planet formation process. Thus, it becomes clear that other existing bodies in our planetary system that have not yet been found abroad should (or at least can) also exist. Two examples are natural satellites (or moons) and trojans. Both types of objects abound in our solar system, in which gas giants host tens of moons and Jupiter has thousands of trojans at both Lagrangian points. Several groups are carrying out the hunt for exomoons, including the Hunt for Exomoons with  Kepler project \citep[HEK,][]{kipping12} and various other works \citep[see, e.g.,][]{weidner10,heller14}. In this project, we deal with the challenge of detecting and characterizing the possible existence of trojans co-orbiting extrasolar planets (hereafter exotrojans). 

The detection of exotrojan bodies is relevant in several aspects. Since these objects are by-products of the planet formation and early evolution processes, they are fossils of the first stages of the life of planetary systems. Thus, they contain primordial dynamical, physical, and chemical information. {For instance, the properties of trojan bodies (e.g., their inclination or libration amplitude in tadpole orbits) and even their mere presence or absence depend on their formation mechanism and can thus be a proof of planet migration; these properties can even discriminate among the different migration mechanisms (see, e.g., \citealt{beauge07}, \citealt{cresswell09}). In the solar system, for example, \cite{morbidelli05} and \cite{nesvorny13} explained the wide variety of properties of the current population of Jupiter trojans as a proof of the dynamical evolution of the gas giant.} Additionally, in the search for habitable worlds, trojans in stable orbits co-revolving with gas giants in the habitable zone of their parent star\footnote{The large majority of planets known to be in the habitable zone of their parent stars are gas giants and so they are not habitable. However, co-orbiting rocky trojans could potentially host liquid water in those cases.} are potential new targets. \cite{dvorak04} investigated the region of habitability for such worlds and concluded that a region around the Lagrangian points of a gas giant in the habitable zone exists where the trojan planet could also be habitable. Thus, although exotic, this possibility should not be disregarded. 

Several techniques have been proposed to detect these bodies; these involve transit timing variations \citep[e.g.,][]{schwarz16,ford07,vokrouhlicky14,haghighipour13}, transits \citep[e.g., ][]{janson13,hippke15}, and radial velocity \citep[][]{ford06,leleu15}. These techniques have been used in previous attempts to detect these bodies, especially with Kepler data \citep[e.g., ][]{hippke15, janson13} and archival radial velocity \citep[e.g.,][]{ford07,madhusudhan08}. However, although some candidates have been proposed (e.g., Kepler-91b by \citealt{lillo-box14} and \citealt{placek15} or KOI-103 by \citealt{janson13}), no exotrojan has been confirmed so far. {Nevertheless, these works have already pointed out that the presence of trojans might explain some observational features (such as the small dims in L$_4$ and L$_5$ found in the combined Kepler light curves by \citealt{hippke15}, Kepler-91 by \citealt{lillo-box14}, {and the case of KIC\,8462852 by \citealt{boyajian16} and \citealt{ballesteros17}}). }

Also, the current knowledge about dynamical stability in these systems allows Earth-size planets to co-orbit with more massive giants, although their formation/capture has yet to be theoretically demonstrated; the largest trojans in our solar system are just few hundreds of kilometers long. If such large bodies exist co-orbiting other planets, their observational imprints should be detectable with future instrumentation such as ESPRESSO \citep{pepe10} or  PLATO \citep{rauer14}. The next generation of precise photometers and high-resolution spectrographs will then push the detection of trojans down to sub-Earth sizes/masses. Here, we take advantage of the current available data from high precision radial velocity instruments on a sample of close-in hot Jupiters to look for trojan planets in the mini-Neptune mass regime ($m_t>10~M_{\oplus}$)\footnote{This limit is due to the fact that archival radial velocity data provide non-dedicated observations for these kinds of searches and because in most cases only data with precisions of  around 10 m/s are available.}.

In this paper, we summarize our current knowledge about trojan bodies  regarding stability regions and orbital dynamics (\S~\ref{sec:stability}), solar system trojans (\S~\ref{sec:SolarSystem}), formation theories,  and their implication in planet formation mechanisms (\S~\ref{sec:formation}).  In  \S~\ref{sec:troy}, we present the goals of the \troy\ project, and the first results from archival radial velocity are then presented in \S~\ref{sec:results}. Finally, we discuss the main results of the paper and provide the conclusions in sections \S~\ref{sec:discussion} and \S~\ref{sec:conclusions}, respectively.

\section{Basics of co-orbital systems: Definitions, naming conventions, and adopted assumptions}
\label{sec:trojans}

\subsection{Dynamics and stability of the Lagrangian points}
\label{sec:stability}

In 1767, Euler found periodic solutions of the three-body problem for which the three bodies are  permanently aligned. The configurations correspond, in the case of the restricted problem, to the equilibrium points called $L_1$, $L_2$, and $L_3$. Lagrange, in 1772,  found two additional periodic solutions, where the three bodies are located at the vertices of an equilateral triangle ($L_4$ and $L_5$ in the restricted problem).

\begin{figure*}[htbp]
\centering 
\includegraphics[width=1.0\textwidth]{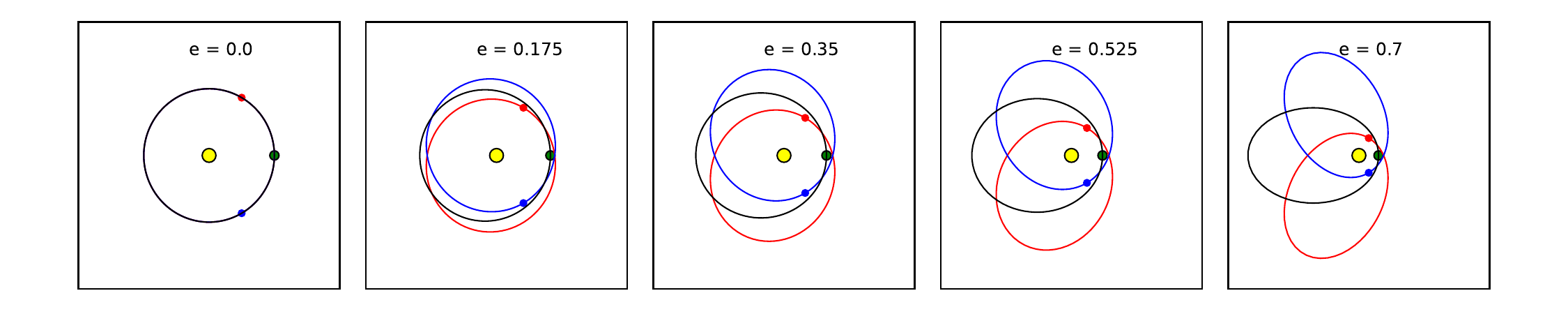} 
\caption{Location of the Lagrangian points L4/L5 for different planet eccentricities (increasing from left to right). The yellow circle represents the star, the black line represents the orbit of the planet, and the blue and red lines represent the orbital path traced by L4 and L5, respectively.}
\label{fig:LP_locations}
\end{figure*}

In the case of eccentric orbits, the orbital path of these Lagrangian points no longer co-rotates in the same orbit as the planet. Generalizations to elliptic Lagrange configurations can be found in \cite{danby64}, \cite{bennett65}, and \cite{roberts02}. Instead, they describe another eccentric orbit with the major axis rotated by $\pm60^{\circ}$ from the direction of the major axis of the planet, always preserving the equilateral triangle but now varying its size at each orbital position. In other words, if we call $\omega_p$ to the argument of the periastron of the orbit of the planet around the star, then the argument of the periastron of the orbit of the trojan body would be $\omega_t = \omega_p \pm 60^{\circ}$, with the "+" symbol for L$_4$ and the "-" for L$_5$ (see Fig.~\ref{fig:LP_locations} for examples of different configurations). \\ 

While the Euler configurations (L$_1$, L$_2$, L$_3$) are unstable \citep{Liouville1842}, the stability of the Lagrange configurations (L$_4$ and L$_5$) depends on the masses of the three bodies, the star ($m_{\star}$),  planet ($m_p$), and trojan ($m_t$).  \cite{gascheau1843} showed that the equilibria are linearly stable as long as 
~
\begin{equation}
\frac{m_p m_t + m_t m_{\star}+ m_p m_{\star}}{(m_t + m_p + m_{\star})^2 } < \frac{1}{27}
\label{eq:gasch}
.\end{equation}
~
If the mass of the planet and the trojan are comparably smaller with the stellar mass (i.e., $m_t<m_p<<m_{\star}$) then this equation can be simplified as $(m_p+m_t)/m_{\star}<1/27$. This limits the stability of the system to configurations in which the sum of the trojan and planet mass is smaller than around 3.7\% of the stellar mass. {Indeed, \cite{sicardy10} demonstrated that this limit can be slightly increased under certain conditions. The stability is also not lost in the case of eccentric orbits (see, for instance, \citealt{danby64,bennett65,roberts02}).} Consequently, the constraint for the stability of co-orbital systems is not strong and allows many configurations (including similar mass planets).
For the unstable cases, \cite{sicardy10} found that a particle left motionless at $L_4$ or $L_5$ quickly escapes from these points and always experiences a close encounter with the planet (see their figure 5). Interestingly, this is one of the proposed mechanisms for the formation of the Moon in the great impactor hypothesis \citep{hartmann75,cameron76}. 
\cite{belbruno05} proposed that \textit{Theia} (a planet embryo that might have impacted the proto-Earth) could have come from one of the Lagrangian points of the Earth.

Since  $L_4$ and $L_5$ are stable against small perturbations, additional bodies can librate around them in stable orbits. For each of these equilibrium points, the three bodies would orbit with the same mean motion around the center of mass of the system. In the planetary case, where $m_p$ and $m_t$ are small with respect to $ m_{\star}$, we call any configuration in which the two planets orbit with the same mean motion around the star \textit{co-orbital} configuration
or 1:1 mean motion resonance (MMR). 

There are different architectures in which two planetary-mass bodies can lie in a 1:1 MMR. \cite{wolf1906} discovered the first co-orbital body in our solar system: Jupiter's trojan Achilles. This was the first located in the vicinity of the $L_4$ and $L_5$ equilibrium points of Jupiter and so the first case of a body in a tadpole configuration, where the bodies librate in the vicinity of the vertices of an equilateral triangle. Later on, using perturbative approaches, \cite{Garfinkel1976,Garfinkel1978} and \cite{erdi1977} modeled the circular co-planar co-orbital resonance in the restricted case ($m_{\star} \gg m_p > m_t=0$). These authors found that, in addition to the tadpole configuration, there is a domain where the resonant angle $\zeta=\lambda_1-\lambda_2$ librates with a large amplitude, while the orbit encompasses the three equilibriums $L_3$, $L_4$, and $L_5$. This is the so-called horseshoe configuration. The first bodies discovered on a horseshoe orbit were the {Saturn satellites Janus and Epimetheus}, found by \cite{SmiReFoLa1980} and \cite{SyPeSmiMo1981}.

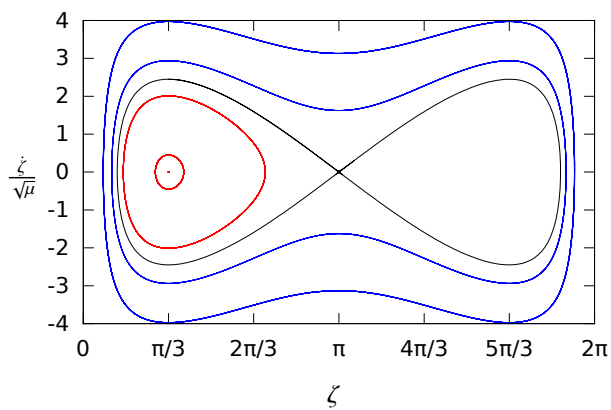
\begin{figure}[h!]
\begin{center}
\includegraphics[width=0.9\linewidth]{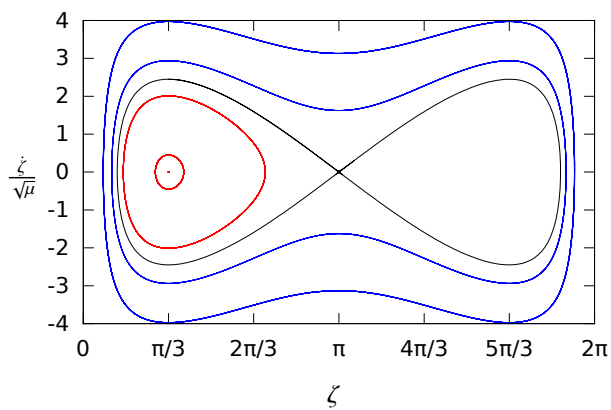}\\
  \setlength{\unitlength}{0.1\linewidth}
\begin{picture}(.001,0.001)
\put(-4.7,3.2){{$\frac{\dot{\zeta}}{\sqrt{\mu}}$}}
\put(0,0){{$\zeta$}}
\end{picture}
\caption{\label{fig:ppH0b}  Phase portrait of equation (\ref{eq:eqerdi}). A separatrix emanating from $L_3$ (black curve) splits the phase space in two different domains: inside the separatrix the region associated with the tadpole orbits (in red) and the horseshoe domain (blue orbits) outside.}
\end{center}
\end{figure}

Recently, \cite{RoPo2013} extended the restricted-case model of Garfinkel and Erdi to the planetary case ($m_{\star} \gg m_p \geq m_t$). Noting $\zeta=\lambda_1-\lambda_2$ the difference of the mean longitudes of the two co-orbitals, $\mu=(m_p+m_t)/(m_p+m_t+m_{\star})$, and $n=2\pi/P$ the mean motion, the equation governing the evolution of $\zeta$ follows the second order differential equation provided by \cite{morais99} in the restricted three-body problem, i.e.,
~
\begin{equation}
\ddot{\zeta} = -3\mu n^2 \left[ 1-\left(2-2\cos{\zeta}\right)^{-3/2}\right]\sin{\zeta}
\label{eq:eqerdi}
.\end{equation}
~
This equation controls the type of motion of the trojan around the Lagrangian points. Its phase portrait is plotted in Fig.~\ref{fig:ppH0b} and possesses the same features as in the restricted case: tadpole orbit (red) librating around $L_4$ and L$_5$ and horseshoe orbits (blue) outside the separatrix emanating from $L_3$.

The stability criterion introduced by Gascheau (Eq.~\ref{eq:gasch}) determines the stability in the immediate vicinity of the L$_4/$L$_5$ equilibriums, hence for $\zeta=\pm60^\circ$. The stability domain around $L_4$ and $L_5$ increases as the quantity $\mu$ decreases, allowing orbits to librate with larger variation of the angle $\zeta$, until stable horseshoe configurations appear for $\mu \approx 3\times 10^{-4}$ or less (Laughlin and Chambers 2001). The stability of the co-orbital configuration depends mainly on the sum of the planetary masses and not as much on the mass repartition between them.

For low inclination and/or eccentricities, tadpole and horseshoe orbits remain the only possible co-orbital configurations, and Eq.~\ref{eq:eqerdi} holds. However, as these parameters increase, new configurations appear, such as quasi-satellites \citep{Namouni1999,MiIn2006} in the eccentric case and retrograde co-orbitals \citep{MoNa2013} in the inclined case. Those configurations, however, are not considered in the present work. Arguably, one can also consider exomoons or binary planets as co-orbitals. But, even though some of the methods we developed might also be adapted to these configurations, they are not a major focus of this project and other teams are already {exploring these possibilities} (e.g., \citealt{kipping12}).

In this work, we use the term trojan and co-orbital indistinguishably to refer to any mass lying or librating around the Lagrangian points of a known planet (either in tadpole or horseshoe orbit). In the case of multiple planets in the system, we consider the Lagrangian points as coming from the gravitational potential of each individual planet with the star, neglecting the gravitational potential of the other planets unless specified, as occurs with the solar system. 

\subsection{Solar system trojans  as potential benchmarks}
\label{sec:SolarSystem}

The only known examples of trojans to date are those in our solar system. These bodies have been discovered in the Lagrangian points of Venus, Mars, Jupiter, Neptune, and recently the Earth. The largest detected accumulation of bodies corresponds to those co-orbiting with Jupiter. The number of trojans catalogued so far increases to $1.6\times 10^{5}$ for sizes larger than 1 km \citep{jewitt00}. The distribution of sizes for these bodies ranges from small, meter-size objects to hundreds of kilometers, where \emph{624 Hektor} is the largest trojan found so far \citep{nicholson1961} with an average diameter of $203\pm 3.6$~km \citep{fernandez03}. Interestingly, according to \cite{jewitt00}, there seems to be two distinguishable populations in the size distribution of Jupiter trojans, smaller or larger  than $r_c\approx 30$~km. The authors explained that this difference in the size distribution has different origins; the largest are primordial objects and the smaller are products of collisional shattering between larger bodies \citep[e.g., ][]{shoemaker89}. These authors also estimated a total mass for the current population of Jupiter trojans of $m_t^{\rm Jup} \approx 9\times 10^{-5} M_{\oplus}$, which is roughly 0.7\% of the mass of the Moon and equivalent to a 400 km radius object of the same density.

A much smaller number of trojans have been found in the orbits of Mars and Neptune. In the first case, seven bodies have been confirmed so far, all of which are roughly 1km or smaller in size \citep{trilling07} and most of which librate around L$_5$, i.e., trailing the planet. In the case of Neptune, 12 trojans have been detected so far according to the Minor Planet Center\footnote{http://www.minorplanetcenter.org/iau/lists/NeptuneTrojans.html}. Their sizes range between 50-200 km and their inclinations range from 1-30 degrees, pointing to a capture origin in contrast to an \emph{in situ} formation (see section~\ref{sec:formation}). However, as stated in \cite{sheppard06}, the population of Neptune trojans is expected to be 20 times larger than that of Jupiter.  Finally, a recent discovery by \cite{connors11} detected the first trojan body co-orbiting in Earth's orbit, 2010 TK$_7$. 

All these solar system discoveries provide information about the expected properties of trojans in planetary systems. In general, the solar system trojans are small ($<$~300~km) and librate with a wide variety of inclinations around the Lagrangian points ($i\in [0^{\circ},50^{\circ}]$) in orbits with moderate eccentricities ($e<0.2$). However, exoplanetary discoveries have shown that we should be ready to encounter the unexpected.

\subsection{Formation theories and implications}
\label{sec:formation}

There are two main scenarios that can lead to the presence of trojan bodies in the Lagrangian points of a planet-star system. First, they could have formed {in situ, potentially being} remnants of the protoplanetary disk trapped in the gravitationally stable regions. The multiple inelastic collisions between dust particles in the first stages (similar to the core accretion process) could have formed larger bodies. 
However, the growth of these planets by collisions of the dust particles and pebbles in orbits around the L$_4$/L$_5$ points is still an open question that continues to puzzle theoreticians {\citep[see ][]{beauge07}.}

Second, they could have been captured during the planet migration along the disk in the first stages of its formation \citep[see, e.g.,][]{namouni17}. It has been suggested and studied from the known population of extrasolar planets (mainly hot and warm Jupiters) that gas giants are formed in the outer parts of the protoplanetary disk and then migrated inward by different mechanisms. In this case, large bodies from the more internal parts of the disk, where rocky planets are expected to grow via the core accretion mechanism, could have been captured in the Lagrangian points of the more massive gas giants. 

{Consequently, given the different nature of each of these formation mechanisms and their relation with the formation of the hosting planet, the resulting trojans would have different orbital and physical properties in each case. Hence, their detection and the characterization of their orbits can provide information about their history and, additionally, about the evolution of the planetary system. For instance, as we mentioned before, the wide variety of properties of the Jupiter trojans was explained by \cite{morbidelli05}, who argued that the population of Jupiter trojan bodies was renewed during the migration of Jupiter and Saturn when they crossed the 1:2 resonance.}


\section{The \troysection\ project}
\label{sec:troy}

As shown in the previous sections, many questions about trojans are open. The aim of the \troy\ project\footnote{\url{http://www.sc.eso.org/~jlillobo/troy/index.html}} is to start a comprehensive and intensive search for co-orbital bodies to known extrasolar planets with various observational techniques. Additionally, we want to understand different theoretical aspects not yet understood about the formation of these bodies. 

We seek to understand how the trojan planets are formed, whether they are captured during the planet migration or if they form in situ in a bottom-up process by collisions of minor bodies. We also want to investigate, in the case of capture (\textit{in situ} formation) scenario, what is the maximum mass that can be trapped (accreted) in the gravity wells of the two bodies while maintaining the stability of the system. We also know from theoretical analysis that planet-like trojans are stable under very relaxed conditions, but is it really possible to grow a planet-like trojan? In other words, is the stability of the Lagrangian points sufficient for allowing the growth of a massive (Earth to super-Earth mass) body? How does this depend on the properties of the planet and/or star? Is there a maximum size for a trojan?

We also aim to understand how common is the existence of trojan planets in extrasolar systems (i.e., estimating their occurrence rate or $\eta-$trojan) and if the relatively small size and mass of the Jupiter's trojans a rule or an exception. In case large bodies can be formed at co-orbital configurations, we are interested to investigate how stable would that planet-like trojan be against minor impacts and migration processes (both smooth disk driven, and more violent dynamical interactions). Also, in the case of a massive trojan, what kind of orbital librations do we expect and how do libration properties (amplitude, inclination, and eccentricity) depend on the trojan mass? Finally, one of the most relevant questions is how the properties of the co-orbital planets can help us to understand the history of the planetary system.

In order to answer these questions, several techniques should be used to look for co-orbital systems around known extrasolar planets. In the following section, we present the first results of a search of trojans around a selected sample of transiting extrasolar planets using archival radial velocity data. We first present the equations derived for the general case of non-circular orbit and then explain the target selection, model fitting, and results.



\section{First results from archival radial velocity}
\label{sec:results}

The reflex motion of a star hosting co-orbitals is, at first approximation, the same as the motion of the star hosting a single planet on a Keplerian orbit because both co-orbitals have the same mean motion. However, if the mass ratio between the two co-orbitals is not too small, the libration amplitude around the Lagrangian equilibrium is not too small and the RVs are precise enough and on a time span that is long enough to observe the libration of the co-orbitals, then one can observe the signature of a co-orbital system in the RV data: the modulation of the amplitude of the signal \citep{laughlin02,leleu15}. 

When this is not the case and the RV data are consistent with a single planet on a Keplerian orbit, a combination of the RV measurement and the information of the time of transit of a planet might solve the degeneracy between a single planet and two co-orbitals; as noticed by \cite{ford06}, if the RV of a star is induced by a pair of co-orbital planets, the predicted time of transit from the RV data is shifted with respect to the actual time of transit of either of the two co-orbitals. Indeed, even though the Keplerian signal in the RV of the star is induced by the barycenter of the two co-orbitals, the time of transit predicted from the RV is hence the time of transit of that barycenter, while the actual planets transit before and after, if at all.

This detection technique was applied by \cite{ford06} to a handful number of known planets at the time, while assuming circular orbits, to set upper limits to the masses of possible trojan bodies in those systems. Also, \cite{madhusudhan08} applied this technique to 25 known planets and found no evidence for a trojan up to their upper mass limits.

A downside of this method is that it is very dependent on the eccentricity of the transiting planet. Any error in the determination of the eccentricity would also produce a shift between the predicted transit time from the RV of a single planet and its actual time of transit. We tackled this problem in a separate paper by \cite{leleu17}, where we generalized the equations presented in \cite{ford06} for the case of eccentric orbits in order to extract as much information as possible from the RV signal. The constraint on the mass of the co-orbital companion to the transiting planet are in any case greatly improved if the secondary eclipse of the transiting planet can be observed, thanks to the precise determination of the parameter $e \cos \omega$.

In this section we explicitly provide the equations used for non-circular orbits (Sect.~\ref{sec:RVeq}) and test these equations with a selected sample of 46 known confirmed transiting exoplanets (Sect.~\ref{sec:targets}) using archival precise radial velocity data from different sources. These data are modeled based on these equations to provide upper limits to the presence of trojans and hints about the possible detection of super-Earth-mass trojans in a few cases.

\subsection{Radial velocity signal from co-orbital planets}
\label{sec:RVeq}

The radial velocity of a star induced by a single planet $k$ on a Keplerian orbit {in the reference frame shown by Fig.~1 in \cite{leleu17}} is written as

   \begin{equation}
v_{k} =  -K_k \left[ \cos{(\nu+\omega_k)} + e_k\cos{\omega_k}  \right] 
   ,\end{equation}

\noindent where $\nu$ is the true anomaly of the planet at time $t$, $e_k$, and $\omega_k$ represent the eccentricity and argument of the periastron of the planet, and $K_k$ is the semi-amplitude of the induced radial velocity of the star. At first order in eccentricity, this equation can be writen as \citep{leleu17}
   \begin{equation}
   \begin{aligned}
v_{k} =  & A_{k} \cos n_k t+ B_{k} \sin n_k t\\
& + C_{k} \cos 2 n_k t + D_{k} \sin 2 n_k t \ ,
\end{aligned}  
   \label{eq:RVeq}
   \end{equation}
with               
   \begin{equation}
      \begin{aligned}
         A_{k} &=- K_{k} \cos\,\varphi_{k} \, ,  &C_{k} &=- e_{k}K_k \cos\, (2\varphi_{k}-\omega_{k}) \, ,\\
        B_{k}&= K_{k} \sin\,\varphi_{k}\, ,   &D_{k}&= e_{k}K_{k} \sin\, (2\varphi_{k}-\omega_{k}) \,  , \\
\end{aligned}
   \label{eq:RV1val}
   \end{equation}   
   and
     \begin{equation}
K_k = \frac{m_k}{M} n_k a_k \sin I_k \ ,\end{equation}

where $I_k$ is the inclination angle between the plane of the sky and the orbital plane, $a_k$ is the semi-major axis, $n_k$ is the mean motion, and $\varphi_k$ an arbitrary phase that depend on the position of the planet at $t=0$.

If we can distinguish the effect of the libration in the RV signal, we can identify co-orbitals from radial velocity alone \citep[see][for the quasi-circular case]{leleu15}. If not, the assumption $n_1=n_2=n$ holds and the mean longitudes of the co-orbitals simply read
\begin{equation}
\lambda_k = n t + \varphi_k + \gO(\mu^{1/2},  e_k^2) \, . \label{eq:longitude}
\end{equation}
For the radial velocity induced by two co-orbitals, we hence sum cosines that have the same frequency. At order one in the eccentricities, we obtain an expression that is equivalent to (\ref{eq:RVeq})
   \begin{equation}
   \begin{aligned}
v  =   & + A \cos n t+ B \sin n t\\
& + C \cos 2 n t + D \sin 2 n t \ ,
\end{aligned}  
   \label{eq:RVeqex}
   \end{equation}
with $A =A_1+A_2$, and similar expressions for $B$, $C,$ and $D$. However, we can differentiate between a single planet and a pair of co-orbitals if we add the constraint of the time of mid-transit. Following Leleu et al. (2017), we set $t=0$ at the time of mid-transit, and we fit the function 
\begin{equation}
\begin{aligned}
v (t) =  \gamma + K \big[&(\alpha -2c) \cos n t  - \sin n t \\
       &+ c \cos 2 n t + d \sin 2 n t \big] \, 
\end{aligned}  
   \label{eq:RVf}
   \end{equation}
to the radial velocity data, where $\gamma$ is the velocity of the center of mass of the system. 

In the case where the transiting planet $m_1$ is alone on its orbit, $K$ is given by the Eq.~\ref{eq:RV1val}, $c=e_1 \cos \omega_1$, $d=e_1 \sin \omega_1$, and $\alpha=0$. However, if the RV are induced by a pair of co-orbitals, $\alpha$ is different from $0$. In the case where $m_2 \ll m_1$, its expression simplifies as
\begin{equation}
\begin{aligned}
    \alpha & = - \frac{m_2}{m_1} \sin \zeta + \gO\left(\left(\frac{m_2}{m_1}\right)^2,e_k^2,\frac{m_2}{m_1} e_k\right)\, .
\end{aligned}  
 \label{eq:alphalim}
   \end{equation}
If $\alpha$ is significantly different from $0$, the system is hence a strong candidate to harbor co-orbitals provided that false positives can be discarded \citep[][]{leleu17}. And inversely, if $\alpha$ is consistent with $0$, an upper mass limit can be inferred for a potential co-orbital companion. {It is important to note at this point that other physical effects can also produce nonzero $\alpha$. These sources of false positives are discussed in section 6 of \cite{leleu17}.}

The expression (\ref{eq:RVf}) shows the importance of the determination of the parameter $c=e_1 \cos \omega_1$ for the determination of $\alpha$. The sensibility to this method depends either on the precision of the radial velocity to determine the $2n$ harmonics in the RV data or on the measurement of the time of secondary eclipse to directly constrain the quantity $c=e_1 \cos \omega_1$.

\subsection{Target selection and data retrieval}
\label{sec:targets}

We applied this technique to already known planets detected by the transit method and that have precise RV measurements. The aim is to constrain the presence of non-negligible masses at their Lagrangian points. The selection of the targets was carried out by imposing the following criteria: i) confirmed planets with measured masses and periods shorter than five~days; ii) the planet must transit so that we have an additional constraint in the RV fitting; iii) the radial velocity data should be precise enough that $m_p/\sigma_{m_p}> 6$ so that the trojan is detectable based on \cite{leleu15}; iv) the estimated RV signal of a 10$M_{\oplus}$ trojan must be above 10 m/s based on the orbital period and stellar mass; and v) there should not be any other known planet in the system, so that no other significant perturbers can affect our calculations. 

Applying these criteria to the current population of known confirmed planets, we selected 55 systems. From these, we removed those systems with a large mass ($> 40 M_{\oplus}$) for the orbital companion, which would be out of the stability criterion for any trojan mass given the masses of the host star and the companion. We also removed S-type planetary systems (i.e., planets orbiting one of the components of a binary system). The final sample is thus composed of 46 single planetary systems, whose main properties from the literature (obtained from exoplanetcatalogue.org) are presented in Table~\ref{tab:literature}.

The radial velocity data for these 46 systems were obtained from different studies on the particular targets. In Table~\ref{tab:rvdata}, we present the number of data points, time span, number of different instruments, and references for each of the studied systems.

\subsection{Modeling}
\label{sec:results}

Based on the equations presented in Sect.~\ref{sec:RVeq}, we have a total of seven parameters to explore: the systemic velocity of the system ($\gamma$), orbital period of the planet ($P_{\rm orb}$), time of mid-transit of the planet ($T_0$), and combined radial velocity semi-amplitude ($K$), $\alpha$, $c$, and $d$. As we saw in section \ref{sec:RVeq}, $c \approx e\cos{\omega}$ at first order in eccentricity, when the contribution of the trojan to the radial velocity is much smaller than that of the planet. Hence, it can be constrained by reported values in the literature from the detection of the secondary eclipse of the planet. In Table~\ref{tab:ecosw}, we provide the values from the literature for $e\cos{\omega}$ for the 20 planets with detected secondary eclipse, among the 46 studied systems. The final prior was adopted as the weighted mean of the detections (when the detection is larger than $3\sigma$) or as null with a $1\sigma$ uncertainty equal to the upper limit when only upper limits could be set. In these cases, we set a Gaussian prior on this parameter with a width equal to five times the estimated uncertainty, $\mathcal{G}(\mu, 5\sigma)$. We set this prior to constrain artificially the convergence through our prior and to allow some freedom in the case of relatively large eccentric values or when we cannot assume that $K_2<<K_1$. In the case of $d$, this parameter can be approximated to $d\approx e\sin{\omega}$ to a first order in eccentricity and $\epsilon$ ($\epsilon=m_t/m_p$). In this case, however, we could not set any constraint to its value and we used a uniform prior in the range $\mathcal{U}(-1,1)$. 

The parameter $\alpha$ is the most relevant in this study since it provides a direct measurement of the radial velocity semi-amplitude induced by the co-orbital planet. A significant deviation from zero would directly indicate the presence of a non-negligible co-orbital mass.  It can be approximated, to a first order in $\epsilon$ to $\alpha\approx \epsilon\sin{\zeta}$, where $\zeta$ is {the angular distance between the two planets} ($\zeta = \pm \pi/3$ at the Lagrangian points L4 and L5) and $\epsilon=K_2/K_1$. Consequently, we can easily see that it is constrained to the range [-1,1] with negative values corresponding to $L_4$ and positive values corresponding to $L_5$. We have used uninformative uniform priors on this value $\mathcal{U}(-1,1)$. 

For the systemic velocity, we assumed a uniform distribution $\mathcal{U}(-100,100)$ km/s. The period and time of mid-transit are very well determined by the transit times. Consequently, we assumed a normal distribution for these parameters with the mean value corresponding to the literature value from the transit analysis and a standard deviation equal to three times the uncertainty in the value, which is $\mathcal{G}(\mu,3\sigma)$. 

Additionally, since the collected radial velocity data of most of the systems come from different instruments and setups, we included $N_{\rm inst}-1$ additional parameters, where $N_{\rm inst}$ is the number of instruments used, to account for the instrumental radial velocity offsets and we also included a jitter for each instrument (i.e., another $N_{\rm inst}$ parameter) to account for random noise from unaccounted instrumentation systematic effects. We used a uniform prior for the offsets of the instrument in the range $\mathcal{U}(-1,1)$~km/s and also for the jitter with a tighter range of $\mathcal{U}(0,0.1)$~km/s.

Additionally the presence of active regions in the stellar surface can give rise to a quasi-periodic RV signal that is modulated by the stellar rotation and active region evolution. In order to model these correlated signals, we used a Gaussian process (GP), which is a nonparametric method that describes the data by evaluating the correlations between each data point. Following previous works with GPs \citep{faria16,rajpaul15,haywood14}, we considered the quasi-periodic kernel

\begin{equation}
\Sigma_{ij} = \eta_1^2 \mathrm{exp}\left[  -\frac{(t_i-t_j)^2}{2\eta_2^2} - \frac{2\sin^2{\frac{\pi (t_i-t_j)}{\eta_3}}}{\eta_4^2}   \right]
,\end{equation}

\noindent where the hyperparameters $\eta_1$, $\eta_2$, $\eta_3$, and $\eta_4$ correspond to the amplitude of the correlations, a timescale for evolution of active regions, a recurrence timescale associated with the stellar rotation period, and a coherence scale for the periodic term, respectively. We used the \textit{george}\footnote{\url{http://dan.iel.fm/george}} package \citep{ambikasaran14} to compute the kernel as a combination of the stationary exponential-squared and the non-stationary exp-sine-squared kernels. The GP likelihood function was used in the MCMC, together with the following priors for the hyperparameters: $\mathcal{LU}(0.1,50)$~m/s, $\mathcal{U}(5,100)$~days, $\mathcal{U}(5,100)$~days, and $\mathcal{LU}(-5,5)$, respectively, where $\mathcal{LU}$ stands for log-uniform prior.

In order to correctly sample the posterior probability distribution of each of those parameters, we used the implementation of Goodman \& Weare's affine invariant Markov chain Monte Carlo (MCMC) ensemble sampler \textit{emcee}\footnote{See http://dan.iel.fm/emcee for further documentation.}, developed by \cite{foreman-mackey13}.  In a first phase, we used 50 walkers and 10000 steps. Then, we resample the position of the walkers in a small N-dimensional ball around the best walker and run a second phase with the same number of walkers and 2500 steps\footnote{This procedure is suggested in the \textit{george} documentation to speed up the convergence.}.
The whole chains of this latter phase are used to compute the marginalized posterior distributions of each parameter  (no thinning was applied). The final chains are thus composed of 1.25$\times 10^5$ steps that are used to compute the marginalized posterior probabilities for each parameter.
\begin{figure*}[htbp]
\centering
\includegraphics[width=1\textwidth]{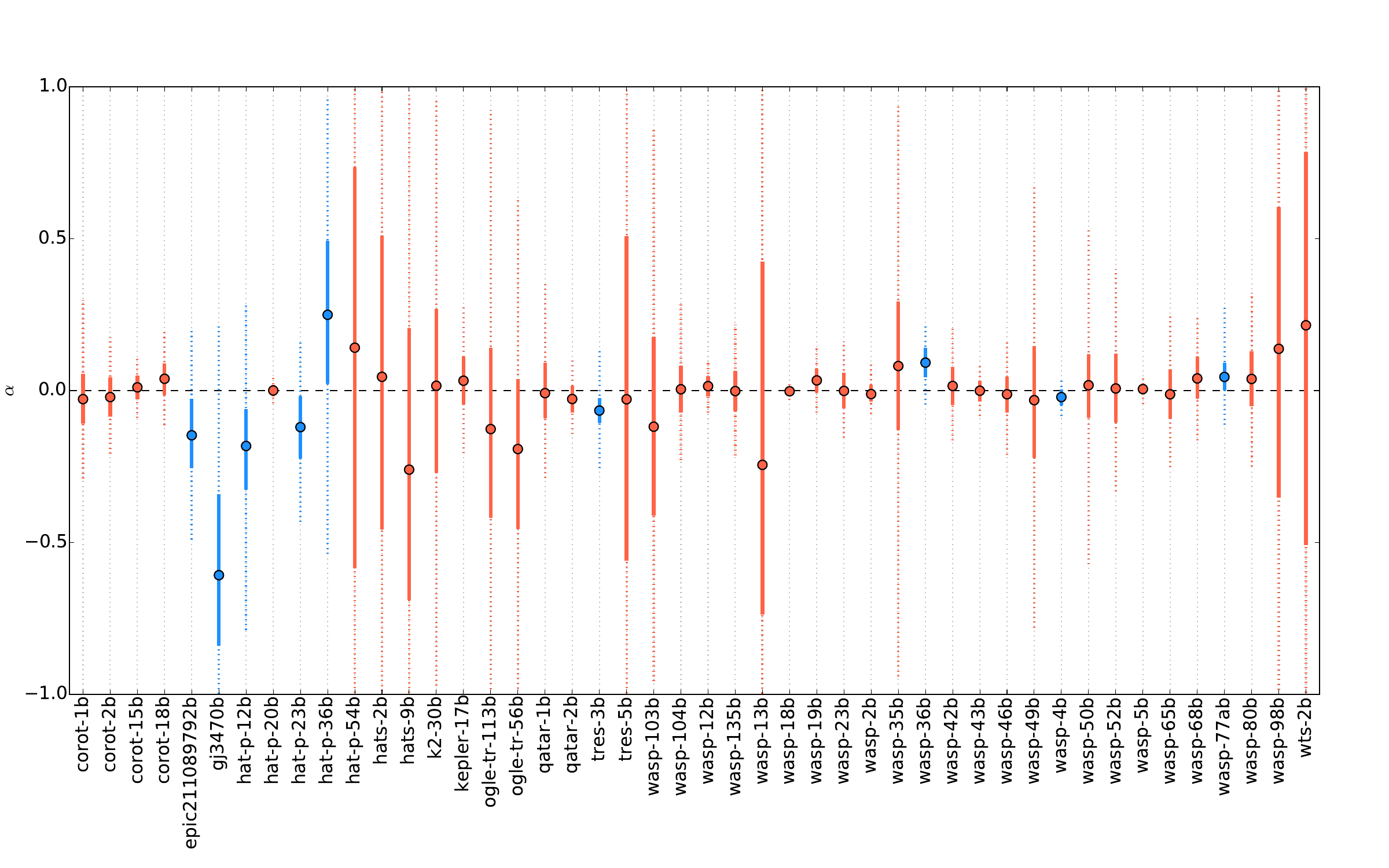}
\caption{Results for the $\alpha$ parameter from the radial velocity analysis. Color error bars indicate the 68.7 confidence intervals (i.e., 1$\sigma$) while the dotted error bars indicate the 99.7\% confidence intervals (3$\sigma$). We show in blue symbols the 9 systems where the null value for $\alpha$ ($\alpha=0$) lies outside of the $1\sigma$ limits.   }
\label{fig:maxmass}
\end{figure*}

\subsection{Results}

The results for the seven orbital and physical parameters (median values from the marginalized posteriors and 68.7\% confidence intervals) are shown in Table~\ref{tab:fiting_results}. Also, the hyperparameters used for the GP together with the instrumentation offsets and jitter values are presented in Table~\ref{tab:hyperparams}. The phase-folded radial velocity plots that correspond to the median model of all chains can be found in Figs.~\ref{fig:rv1}-\ref{fig:rv2}. Here we removed the GP corresponding to the median model,  included 200 models randomly selected among those with all parameters within 3$\sigma,$  and subtracted the GP model.

First of all,  in most cases the GP hyperparameters that account for possible correlated noise from stellar activity are not constrained, where $\eta_1$ is usually below 1 m/s and mostly several times below the RV precision of the instruments used. In practice, this means there is no stellar activity affecting the RV data at the level of the measured RV precision. 

Among all parameters, $\alpha$ is the most relevant parameter, since it determines whether the present data are sufficient to detect a non-negligible mass at the Lagrangian points with significance. In Fig.~\ref{fig:maxmass}, we show the 68.7\% (1$\sigma$) and 99.7\% (3$\sigma$) confidence intervals for this parameter on each system. As shown, in none of the studied systems the $\alpha$ parameter excludes the null value at a 99.7\% confidence (i.e., $3\sigma$). Consequently, we cannot claim the detection of trojans in this sample. However, in 9 cases, this parameter is more than $1\sigma$ away from the null value. And in two of them (WASP-36 and GJ\,3470), the parameter is $>2\sigma$ away. {Hence, no significant detection is found among the studied systems in this work.}

Based on the posterior probability functions of this parameter, assuming $m_t<<m_p$ and to a first order in eccentricity, we can convert the $\alpha$ parameter into a co-orbital mass for the restricted case (i.e., assuming the object is exactly at one of the Lagrangian points, so $m_t \approx -\alpha\, m_p/\sin{\zeta} $, with $\zeta = \pm 60^{\circ}$). As an example, a Jupiter-mass planet with a 10\,$M_{\oplus}$ trojan would have $\alpha \sim 0.036$. Since no significant detections has been found, we provide the 95.4\% upper mass limit for each of the Lagrangian points in Table~\ref{tab:maxmass}. In this table, we also show the $\alpha/\sigma_{\alpha}$ ratios, which provide a quick estimation  of the significance of the detection. 

In some cases, the uncertainty of the  $\alpha$ parameter is relatively large. This is due to a combination of a small dataset and a sparse distribution of the measurements, which prevents us from constraining the GP hyperparameters. On the contrary, we can see that when a large number of data points are available ($>50$) with sufficiently precise data ($\sigma_{RV}\sim 5$ m/s or better) we can start exploring the $<30~M_{\oplus}$ regime in the Lagrangian points. A good example of this is, for instance, WASP-2, where with 56 measurements from 5 different instruments, we are able to discard trojans with masses above 15.3 $M_{\oplus}$  at L$_5$ and 14.6 $M_{\oplus}$ at L$_4$ at a $3\sigma$ level. This demonstrates the importance of accumulating a large number of radial velocity points along the whole orbital phase to narrow the posterior probability of the $\alpha$ parameter. Consequently, subsequent follow-up on the best candidates is absolutely necessary in order to i) accumulate more data points spread along the whole orbital phase and ii) cover the rotational period of the star to estimate the correlated noise. Finally, a detailed analysis of the stability of these systems will be performed in a separate work and is out of the scope of this paper.


\section{Discussion}
\label{sec:discussion}

\subsection{Characteristics of the analysis}

\begin{figure}
\centering
\includegraphics[width=0.48\textwidth]{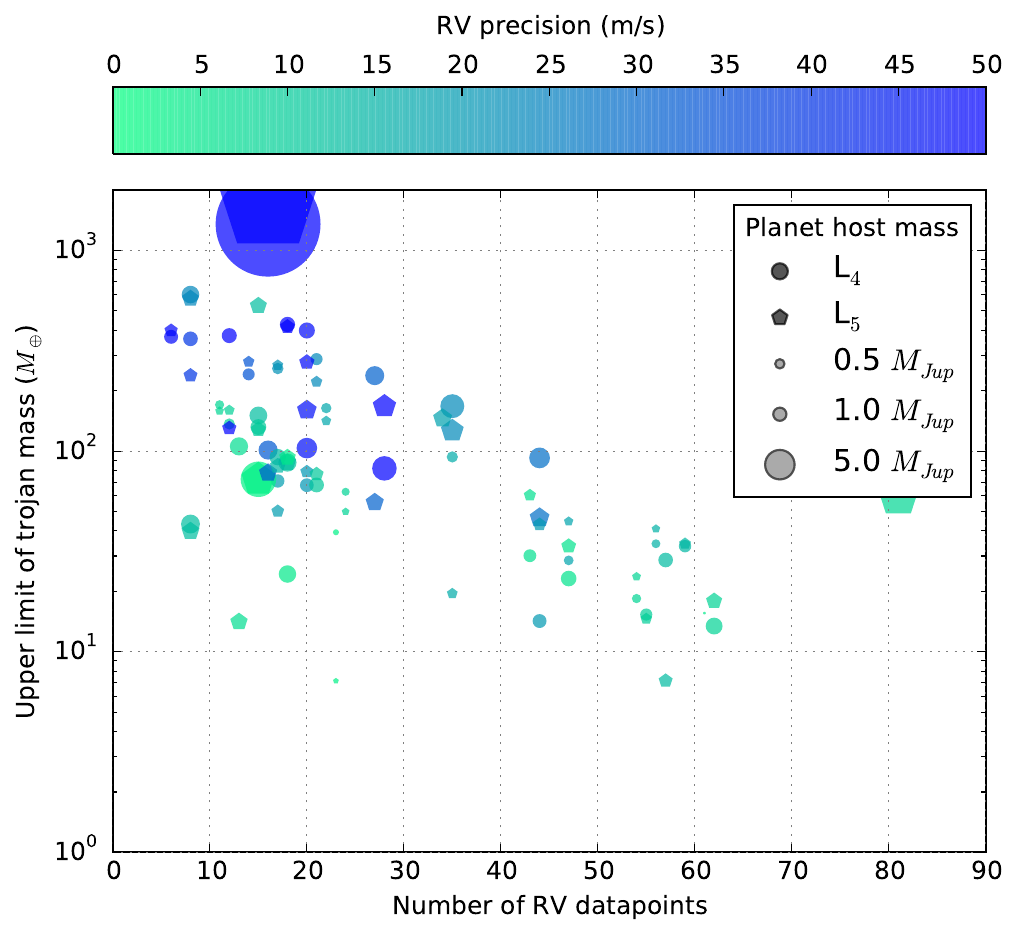}
\caption{Relation between the number of radial velocity points available for the analysis and the upper limit (95.4\% confidence interval) on the trojan mass for each of the 46 systems. In color code we represent the mean precision of all RV measurements. The $\alpha$ 95.4\% confidence interval is transformed into L$_4$ (circles) and L$_5$ (pentagons) mass values assuming $m_t<<m_p$, small eccentricities, and trojans close to the Lagrangian points.}
\label{fig:maxmass_nn}
\end{figure}

The radial velocity analysis described in the previous sections is the most dedicated and statistically robust radial velocity study about the presence of trojan bodies in a relatively large sample of extrasolar systems. But, it is important to highlight that given the selection criteria, we are biased toward short periods ($P<5$~days) and massive planets ($M_p>0.3~M_{\rm Jup}$ and most with $M_p\sim1-2~M_{\rm Jup}$). Dedicated observations with currently operational high-resolution (stabilized) instruments can go beyond these limits, although a relatively large amount of time would be needed.  The results presented in this work demonstrate that current instrumentation is capable of ruling out trojans with masses $>30~M_{\oplus}$ for a minimum number of measurements of $N\sim50$ and a typical radial velocity precision of $\sim10$~m/s. This is clear from Fig.~\ref{fig:maxmass_nn}, where we show the upper mass limits for trojans at L$_4$ and L$_5$ in the 46 systems studied by assuming 95.4\% confident intervals and the trojan located exactly at the Lagrangian points (i.e., $\zeta=\pm60^{\circ}$) depending on the number of data points. The mean upper mass limit, including all systems with more than 50 points, is $30~M_{\oplus}$. The mean precision is  27 m/s, which is worse than the precision of the current state-of-the-art spectrographs) and the number of points is 28. It is relevant to note that, even with archival data, in some cases we are able to provide upper mass limits to trojan masses at one of the Lagrangian points at the order of few Earth masses.

\subsection{Implications for previous works}

\cite{madhusudhan08} performed a similar analysis for 25 systems and they provided a mass sensitivity of $56~M_{\oplus}$. In our work, we used additional data gathered for 46 systems and used a direct fitting of the RV to investigate the presence of trojans at $L_4/L_5$. Given that our method is more general and valid for any value of $\zeta$, in order to provide upper limits to the mass of the trojans and without any other constraint, we need to make an assumption about its value. With this purpose, we hereafter assume the trojans to be located exactly at the Lagrangian points (i.e., $\zeta=\pm 60^{\circ}$). We have six systems in common with \cite{madhusudhan08}, for which we can compare how the larger number of data points decreases the upper limit on the trojan mass. In particular, they (we) find 95.4\% upper limits of 
117 $M_{\oplus}$ (71 $M_{\oplus}$ at L$_4$ and 50 $M_{\oplus}$ at L$_5$) with 9 (17) measurements for CoRoT-1, 
153 $M_{\oplus}$ (167 $M_{\oplus}$ at L$_4$ and 126 $M_{\oplus}$ at L$_5$) with 24 (35) measurements for CoRoT-2,  
199 $M_{\oplus}$ (15.3 $M_{\oplus}$ at L$_4$ and 14.6 $M_{\oplus}$ at L$_5$) with 7 (56) measurements for WASP-2,
43   $M_{\oplus}$ (29 $M_{\oplus}$ at L$_4$ and 7.2 $M_{\oplus}$ at L$_5$) with 13 (57) measurements for WASP-4,
81.3   $M_{\oplus}$ (105 $M_{\oplus}$ at L$_4$ and 14.1 $M_{\oplus}$ at L$_5$) with 11 (13) measurements for TrES-3,
54.7   $M_{\oplus}$ (13.4 $M_{\oplus}$ at L$_4$ and 17.9 $M_{\oplus}$ at L$_5$) with 11 (46) measurements for WASP-5. 
It is clear from these numbers how increasing the number of points in each dataset clearly decreases the upper limit that we can explore (in particular, when $N> 50$). 

Other previously analyzed system in the context of trojan planets is WASP-12. In a recent paper by \cite{kislyakova16}, the authors have sought to explain some intriguing features in the ultraviolet light curve of the hot Jupiter in this system by assuming the presence of Io-like trojans. In particular, in the case of  WASP-12, an early ingress in the ultraviolet transit of the  planet was found as compared to the optical transit times; the egress time agrees in all wavelengths analyzed. Additionally, \cite{fossati10} and \cite{haswell12} found a complete suppression of emission line cores of Mg{\sc{ii}} h\&k  in the near ultraviolet and Ca{\sc{ii}} H\&K in the optical regime. These observables could be explained by absorption in a hypothetical bow shock ahead of the planet. They were investigated by \cite{ben-Jaffel14}, who have proposed that this could be caused by volcanic outgassing of Io-like exomoons creating a plasma tori around the planet. Given the dynamical difficulties of having a large exomoon in such close-in planets, \cite{kislyakova16} have proposed that  this plasma could come from outgassing of lava oceans on the surface of  rocky trojans, what would release Mg and Ca. WASP-12 is one of the targets analyzed in this paper. Our analysis shows no evidence for the presence of trojan planets more massive than 23.2 $M_{\oplus}$ at L$_4$ and 33.6 $M_{\oplus}$ at L$_5$. Testing the presence of Io-like masses ($\sim 0.015 M_{\oplus}$) is unaffordable with current instrumentation. However, this was just a toy model to explain the mentioned features of WASP-12\,b and more massive trojans could equally explain these results. Consequently, this is still a good candidate to continue monitoring.  

{Regarding formation scenarios, \cite{beauge07} performed simulations to estimate the maximum mass that can be aggregated in a tadpole region to form a terrestrial-like planet through accretional collisions of rocky planetesimals; these authors found the maximum mass is $\sim$~0.6 M$_{\oplus}$. Thus, the detection of more massive trojans would be a clear indication that other formation mechanisms (e.g., capture or gas-instability collapse) should play a role. In this paper we have found candidates with masses well above this boundary. Unfortunately, the current precision of the archival data is not sufficient to confirm the detections and so we cannot conclude on their formation scenarios.}

\section{Conclusions}
\label{sec:conclusions}

In this paper we have presented the \troy\ project, whose main aim is to start a detailed, multi-technique, and dedicated search for co-orbital planets. As shown by previous theoretical works, the existence of massive (Earth-mass) trojans in the Lagrangian points of massive planets is allowed under certain conditions, keeping the system stable during planet life timescales. The only limitation now is whether there exists a mechanism to place these bodies in the gravity wells of more massive planets. Theoretically, we still do not have an answer to whether rocky worlds can be formed within the Lagrangian points in a bottom-up process as Earth-like planets do. Also, to our knowledge, no dynamical studies have been carried out about the stability of a captured rocky planet in the Lagrangian point of a gas giant in different events such as migration or additional perturbations (resonant planets, other trojans, etc.).

We started an intensive search by means of different observing techniques using archival and newly acquired data. In this paper we presented the analysis of the archival radial velocity of 46 (apparently) single-planet systems. We used the newly derived equations from \cite{leleu17} to test the presence of non-negligible masses at the Lagrangian points of these planets by means of the $\alpha$ parameter. The detection of a significant deviation of this parameter from the null value directly suggests the presence of a trojan planet; other possible mimicking configurations are discussed in \citealt{leleu17}, where we have concluded that all of these other configurations can be ruled out with other techniques such as TTVs or the same RV data.

The results of this first study have provided upper mass limits for trojans at L$_4$/L$_5$ in all 46 systems given the present data. Interestingly, we have found nine cases where $\alpha$ is at least $1\sigma$ away from the null value, although some of the posteriors are too broad to extract clear conclusions. Furthermore, in two cases (GJ\,3470 and WASP-36), the median value for the $\alpha$ parameter is  $>2\sigma$ away from the null hypothesis. 

\begin{figure}
\centering
\includegraphics[width=0.48\textwidth]{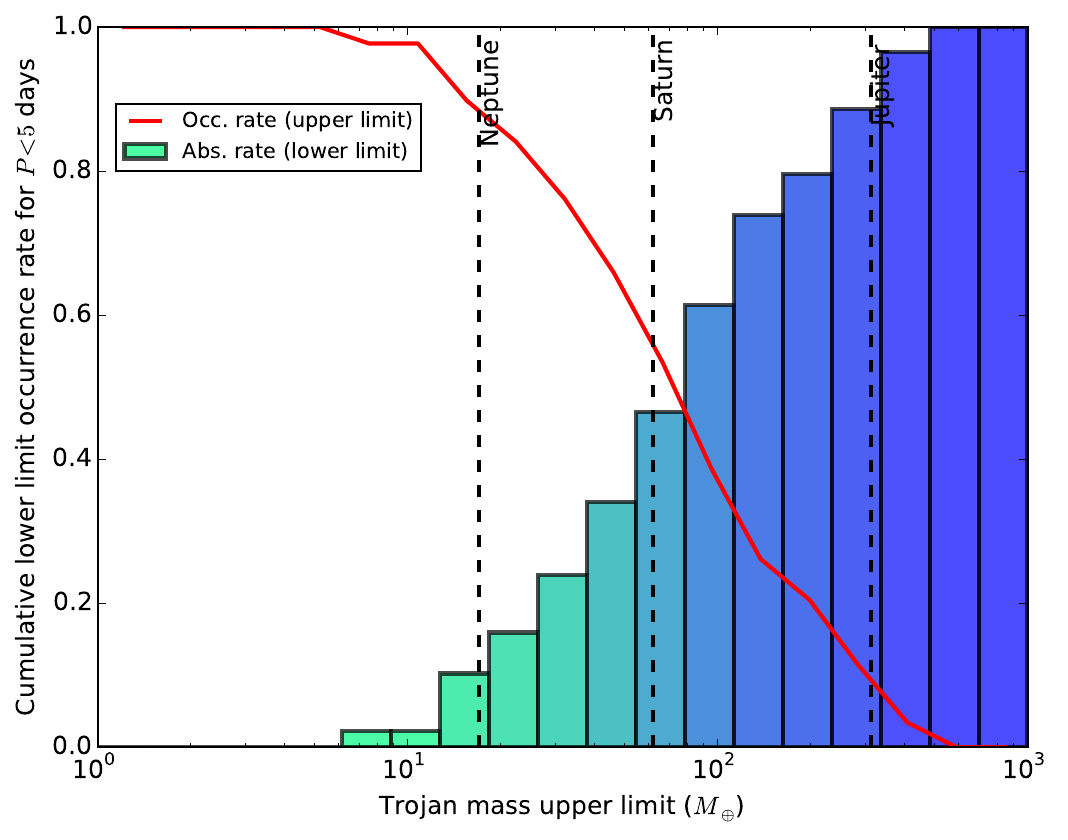}
\caption{Estimation of the upper limits of the occurrence rate of trojans in the particular sample of planetary systems studied in this work. The upper mass limits were calculated as 95.4\% confidence level on the $\alpha$ parameter and translated to mass values by assuming the trojan is located exactly at the Lagrangian points. The histogram shows the lower limit for the absence rate (color coded by the y-axis value), while the red line indicates the upper limit for the occurrence rate (inverted histogram, included for illustration purposes).}
\label{fig:eta_trojan}
\end{figure}

Even though in the low number statistics, given the upper mass limits provided in this paper, we can start estimating occurrence rates of exotrojans in the particular sample studied here (i.e., short-period -$P<5$~days- single planets). In particular, since we only detect upper mass limits, we can estimate the upper limits for the occurrence rate of trojans up to a certain mass, which is defined as the 95.4\% confidence level for the $\alpha$ parameter assuming that the trojan is located exactly at the Lagrangian point. In Fig.~\ref{fig:eta_trojan}, we present these values for the sample studied in this work. According to this, we can say in general terms that at least 12\% of planets with periods shorter than five days do not have co-orbital planets more massive than Neptune. Equivalently, at least 50\% lack trojans more massive than Saturn. Also, we can discard Jupiter-mass trojans in this particular sample at a $\sim$~90\% level. The reasons for this absence of massive trojans can be numerous (e.g., difficulties in forming such large bodies in situ at the Lagrangian points or keeping them stable during planetary lifetimes, difficulties in capturing such massive planets in stable orbits around the Lagrangian points, etc.). But in any case, the evidence presented here for each individual system can inform formation and migration models. New data at higher precision with current instrumentation will certainly improve these estimations and better constrain the presence of lower mass trojans.

This study thus represents the largest radial velocity analysis of the existence of trojan planets in extrasolar systems so far. We finally want to point out that although exotic, the discovery of the wide variety of extrasolar planets and up-to-now hostile environments where they can live, has highlighted the many surprises that nature can bring up. The fact that no large trojans exist in our solar system could just be a hint of a chaotic early evolution during the first stages of its formation and the migration of the gaseous giants Jupiter and Saturn. Whether Earth-size or larger trojans are common or not, or even if they exist or not is still an open question that will be scrutinized by the \troy\ project.


\begin{acknowledgements}
We thank the referee for her/his useful comments during this process. J.L-B acknowledges financial support from the Marie Curie Actions of the European Commission (FP7-COFUND). DB acknowledges financial support from the Spanish grant ESP2015- 65712-C5-1-R. PF and NCS acknowledge support by Funda\c{c}\~ao para a Ci\^encia e a Tecnologia (FCT) through Investigador FCT contracts of reference IF/01037/2013/CP1191/CT0001 and IF/00169/2012/CP0150/CT0002, respectively, and POPH/FSE (EC) by FEDER funding through the program ``Programa Operacional de Factores de Competitividade - COMPETE''. PF further acknowledges support from Funda\c{c}\~ao para a Ci\^encia e a Tecnologia (FCT) in the form of an exploratory project of reference IF/01037/2013/CP1191/CT0001. J.P.F. acknowledges support from FCT through the grant reference SFRH/BD/93848/2013. A.C.M.C acknowledges financial support from the Observatoire de Paris Scientific Council, CIDMA strategic project UID/MAT/04106/2013. P.R. acknowledges financial support from the Programme National de Plan\'etologie (INSU-CNRS).
\end{acknowledgements}

\bibliographystyle{aa} 
\bibliography{../../biblio2} 

\newpage

\begin{figure*}
\centering
\includegraphics[width=1\textwidth]{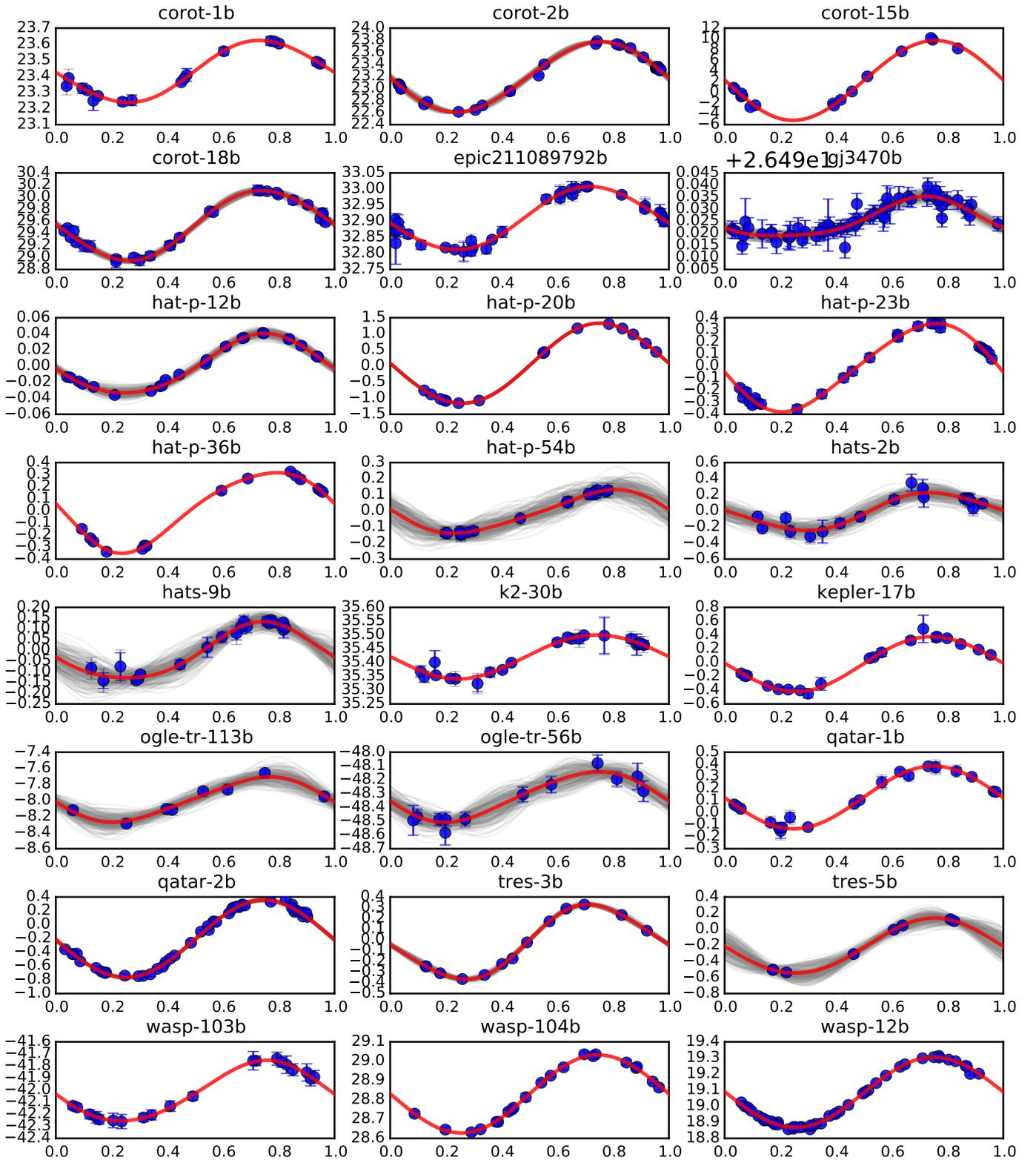}
\caption{Phase-folded radial velocity curves (in km/s) for the 46 systems studied (continuation in Fig.~\ref{fig:rv2}). The blue circles represent the RV data (including errorbars). The red solid line shows the model corresponding to the median of the marginalized posterior distribution for each parameter. The gray lines show 100 models randomly chosen from the final MCMC chain.}
\label{fig:rv1}
\end{figure*}

\begin{figure*}
\centering
\includegraphics[width=1\textwidth]{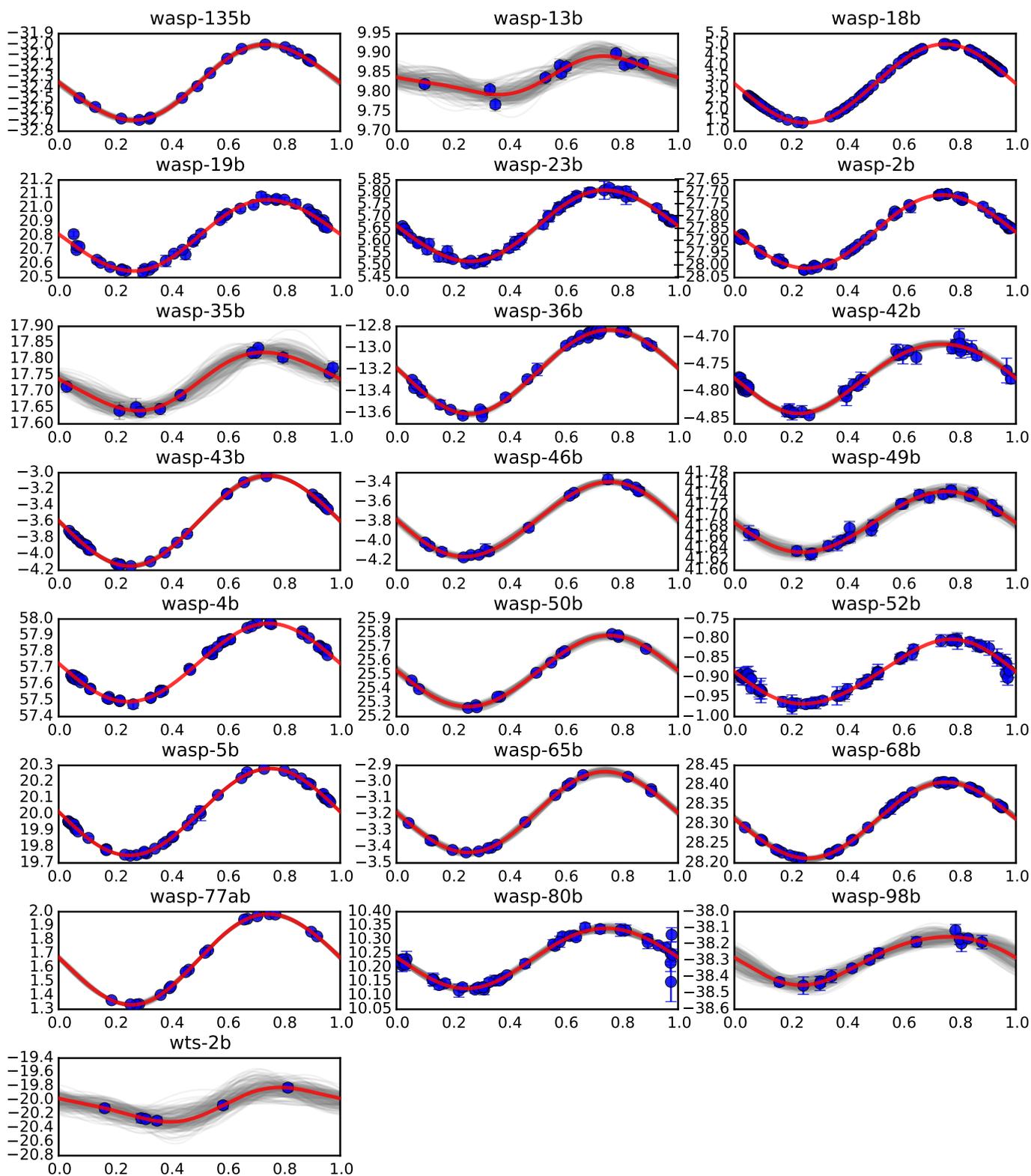}
\caption{Continuation of Fig.~\ref{fig:rv1}. See its caption for details.}
\label{fig:rv2}
\end{figure*}

\newpage

\begin{table*}[]
\setlength{\extrarowheight}{3pt}
\caption{Archival parameters}
\scriptsize
\begin{center}
\begin{tabular}{llllllll}
\hline
\hline

            & Period & $T_0-2400000$ & $e$ & $\omega$ & $i$ & $M_{\star}$ & \\ 
Object & (days) & $(days)$ &  & deg. & deg. & $M_{\odot}$ & \\ \hline

corot-1b             & 	$1.5089557^{+0.0000064}_{-0.0000064}$ & 	$54159.45320^{+0.00010}_{-0.00010}$ & 	- & 	- & 	$85.10^{+0.50}_{-0.50}$ & 	$0.95^{+0.15}_{-0.15}$ & 	 \\ 
corot-2b             & 	$1.7429964^{+0.0000017}_{-0.0000017}$ & 	$54706.4041^{+0.0030}_{-0.0030}$ & 	- & 	- & 	$87.84^{+0.10}_{-0.10}$ & 	$0.970^{+0.060}_{-0.060}$ & 	 \\ 
corot-15b            & 	$3.060360^{+0.000030}_{-0.000030}$ & 	$54753.5608^{+0.0011}_{-0.0011}$ & 	- & 	- & 	$86.7^{+2.3}_{-3.2}$ & 	$1.32^{+0.15}_{-0.15}$ & 	 \\ 
corot-18b            & 	$1.9000693^{+0.0000028}_{-0.0000028}$ & 	$55321.72412^{+0.00018}_{-0.00018}$ & 	- & 	- & 	$86.50^{+0.90}_{-0.90}$ & 	$0.95^{+0.15}_{-0.15}$ & 	 \\ 
epic211089792b       & 	$3.2588321^{+0.0000019}_{-0.0000019}$ & 	$53219.0095^{+0.0022}_{-0.0022}$ & 	$0.066^{+0.022}_{-0.022}$ & 	$132^{+21}_{-21}$ & 	$86.656^{+0.11}_{-0.080}$ & 	$0.940^{+0.020}_{-0.020}$ & 	 \\ 
gj3470b              & 	$3.336710^{+0.000050}_{-0.000050}$ & 	$56090.47690^{+0.00015}_{-0.00015}$ & 	- & 	- & 	$88.30^{+0.50}_{-0.50}$ & 	$0.539^{+0.047}_{-0.047}$ & 	 \\ 
hat-p-12b            & 	$3.2130598^{+0.0000021}_{-0.0000021}$ & 	$54419.19556^{+0.00020}_{-0.00020}$ & 	- & 	- & 	$89.00^{+0.40}_{-0.40}$ & 	$0.730^{+0.020}_{-0.020}$ & 	 \\ 
hat-p-20b            & 	$2.8753170^{+0.0000040}_{-0.0000040}$ & 	$56708.356260^{+0.000088}_{-0.000088}$ & 	$0.01500^{+0.00050}_{-0.00050}$ & 	$(3.2^{+1.3)e+02}_{-1.3)e+02}$ & 	$86.80^{+0.20}_{-0.20}$ & 	$0.756^{+0.028}_{-0.028}$ & 	 \\ 
hat-p-23b            & 	$1.212884^{+0.000020}_{-0.000020}$ & 	$54852.26464^{+0.00018}_{-0.00018}$ & 	$0.106^{+0.044}_{-0.044}$ & 	$118^{+25}_{-25}$ & 	$85.1^{+1.5}_{-1.5}$ & 	$1.130^{+0.050}_{-0.050}$ & 	 \\ 
hat-p-36b            & 	$1.3273470^{+0.0000030}_{-0.0000030}$ & 	$55565.18144^{+0.00020}_{-0.00020}$ & 	$0.063^{+0.032}_{-0.032}$ & 	$95^{+63}_{-63}$ & 	$86.0^{+1.3}_{-1.3}$ & 	$1.022^{+0.049}_{-0.049}$ & 	 \\ 
hat-p-54b            & 	$3.799847^{+0.000014}_{-0.000014}$ & 	$56299.30370^{+0.00024}_{-0.00024}$ & 	- & 	- & 	$87.040^{+0.084}_{-0.084}$ & 	$0.645^{+0.020}_{-0.020}$ & 	 \\ 
hats-2b              & 	$1.3541330^{+0.0000010}_{-0.0000010}$ & 	$55954.585760^{+0.000090}_{-0.000090}$ & 	- & 	- & 	$87.20^{+0.70}_{-0.70}$ & 	$0.882^{+0.037}_{-0.037}$ & 	 \\ 
hats-9b              & 	$1.9153073^{+0.0000052}_{-0.0000052}$ & 	$56124.25896^{+0.00086}_{-0.00086}$ & 	- & 	- & 	$86.5^{+1.6}_{-2.5}$ & 	$1.030^{+0.040}_{-0.040}$ & 	 \\ 
k2-30b               & 	$4.098507^{+0.000028}_{-0.000028}$ & 	$57063.80710^{+0.00027}_{-0.00027}$ & 	$0.027^{+0.036}_{-0.020}$ & 	$120^{+1.0)e+02}_{-51}$ & 	$86.32^{+0.38}_{-0.38}$ & 	$0.984^{+0.020}_{-0.020}$ & 	 \\ 
kepler-17b           & 	$1.48571080^{+0.00000020}_{-0.00000020}$ & 	$55185.678030^{+0.000026}_{-0.000026}$ & 	- & 	- & 	$87.20^{+0.15}_{-0.15}$ & 	$1.160^{+0.060}_{-0.060}$ & 	 \\ 
ogle-tr-113b         & 	$1.4324772^{+0.0000012}_{-0.0000012}$ & 	$53471.77820^{+0.00050}_{-0.00050}$ & 	- & 	- & 	$89.40^{+0.60}_{-0.60}$ & 	$0.780^{+0.020}_{-0.020}$ & 	 \\ 
ogle-tr-56b          & 	$1.2119090^{+0.0000010}_{-0.0000010}$ & 	$53936.5980^{+0.0010}_{-0.0010}$ & 	- & 	- & 	$78.80^{+0.50}_{-0.50}$ & 	$1.170^{+0.040}_{-0.040}$ & 	 \\ 
qatar-1b             & 	$1.42002460^{+0.00000070}_{-0.00000070}$ & 	$56157.42204^{+0.00010}_{-0.00010}$ & 	- & 	- & 	$84.52^{+0.24}_{-0.24}$ & 	$0.850^{+0.030}_{-0.030}$ & 	 \\ 
qatar-2b             & 	$1.3371182^{+0.0000037}_{-0.0000037}$ & 	$55624.26679^{+0.00011}_{-0.00011}$ & 	- & 	- & 	$88.30^{+0.94}_{-0.94}$ & 	$0.740^{+0.037}_{-0.037}$ & 	 \\ 
tres-3b              & 	$1.306^{+0.060}_{-0.070}$ & 	$54538.58069^{+0.00021}_{-0.00021}$ & 	- & 	- & 	$82.15^{+0.21}_{-0.21}$ & 	$0.924^{+0.040}_{-0.040}$ & 	 \\ 
tres-5b              & 	$1.48224460^{+0.00000070}_{-0.00000070}$ & 	$55443.25153^{+0.00011}_{-0.00011}$ & 	- & 	- & 	$84.5290^{+0.0050}_{-0.0050}$ & 	$0.893^{+0.024}_{-0.024}$ & 	 \\ 
wasp-103b            & 	$0.925542^{+0.000019}_{-0.000019}$ & 	$56459.59957^{+0.00075}_{-0.00075}$ & 	- & 	- & 	$86.3^{+2.7}_{-2.7}$ & 	$1.2200^{+0.0039}_{-0.0039}$ & 	 \\ 
wasp-104b            & 	$1.7554137^{+0.0000018}_{-0.0000036}$ & 	$56406.11126^{+0.00012}_{-0.00012}$ & 	- & 	- & 	$83.63^{+0.25}_{-0.25}$ & 	$1.020^{+0.090}_{-0.090}$ & 	 \\ 
wasp-12b             & 	$1.0914222^{+0.0000011}_{-0.0000011}$ & 	$54508.97605^{+0.00028}_{-0.00028}$ & 	- & 	- & 	$86.0^{+3.0}_{-3.0}$ & 	$1.35^{+0.14}_{-0.14}$ & 	 \\ 
wasp-135b            & 	$1.40137940^{+0.00000080}_{-0.00000080}$ & 	$55230.99020^{+0.00090}_{-0.00090}$ & 	- & 	- & 	$82.00^{+0.60}_{-0.60}$ & 	$0.980^{+0.060}_{-0.060}$ & 	 \\ 
wasp-13b             & 	$4.353011^{+0.000013}_{-0.000013}$ & 	$55575.5136^{+0.0016}_{-0.0016}$ & 	- & 	- & 	$85.64^{+0.24}_{-0.24}$ & 	$1.090^{+0.050}_{-0.050}$ & 	 \\ 
wasp-18b             & 	$0.94145180^{+0.00000040}_{-0.00000040}$ & 	$54221.48163^{+0.00038}_{-0.00038}$ & 	$0.0088^{+0.0012}_{-0.0012}$ & 	$269.0^{+3.0}_{-3.0}$ & 	$86.0^{+2.5}_{-2.5}$ & 	$1.240^{+0.040}_{-0.040}$ & 	 \\ 
wasp-19b             & 	$0.78884000^{+0.00000030}_{-0.00000030}$ & 	$55168.968010^{+0.000090}_{-0.000090}$ & 	$0.0046^{+0.0044}_{-0.0044}$ & 	$3^{+70}_{-70}$ & 	$79.40^{+0.40}_{-0.40}$ & 	$0.904^{+0.045}_{-0.045}$ & 	 \\ 
wasp-23b             & 	$2.9444256^{+0.0000011}_{-0.0000013}$ & 	$55320.12363^{+0.00013}_{-0.00013}$ & 	- & 	- & 	$88.39^{+0.79}_{-0.45}$ & 	$0.78^{+0.13}_{-0.13}$ & 	 \\ 
wasp-2b              & 	$2.15222144^{+0.00000040}_{-0.00000040}$ & 	$53991.51530^{+0.00017}_{-0.00017}$ & 	- & 	- & 	$84.73^{+0.19}_{-0.19}$ & 	$0.84^{+0.11}_{-0.11}$ & 	 \\ 
wasp-35b             & 	$3.1615750^{+0.0000020}_{-0.0000020}$ & 	$55531.47907^{+0.00015}_{-0.00015}$ & 	- & 	- & 	$87.96^{+0.25}_{-0.25}$ & 	$1.100^{+0.080}_{-0.080}$ & 	 \\ 
wasp-36b             & 	$1.5373653^{+0.0000027}_{-0.0000027}$ & 	$55569.837310^{+0.000093}_{-0.000093}$ & 	- & 	- & 	$83.65^{+0.22}_{-0.22}$ & 	$1.020^{+0.032}_{-0.032}$ & 	 \\ 
wasp-42b             & 	$4.9816872^{+0.0000073}_{-0.0000073}$ & 	$55650.56720^{+0.00023}_{-0.00023}$ & 	$0.060^{+0.013}_{-0.013}$ & 	$167^{+26}_{-26}$ & 	$88.25^{+0.27}_{-0.27}$ & 	$0.890^{+0.080}_{-0.080}$ & 	 \\ 
wasp-43b             & 	$0.81347753^{+0.00000070}_{-0.00000070}$ & 	$55726.54336^{+0.00012}_{-0.00012}$ & 	$0.0035^{+0.0025}_{-0.0025}$ & 	$328^{+34}_{-34}$ & 	$82.33^{+0.20}_{-0.20}$ & 	$0.717^{+0.025}_{-0.025}$ & 	 \\ 
wasp-46b             & 	$1.4303700^{+0.0000023}_{-0.0000023}$ & 	$55392.31553^{+0.00020}_{-0.00020}$ & 	- & 	- & 	$82.63^{+0.38}_{-0.38}$ & 	$0.956^{+0.034}_{-0.034}$ & 	 \\ 
wasp-49b             & 	$2.7817387^{+0.0000056}_{-0.0000056}$ & 	$55580.59436^{+0.00029}_{-0.00029}$ & 	- & 	- & 	$84.89^{+0.19}_{-0.19}$ & 	$0.940^{+0.070}_{-0.070}$ & 	 \\ 
wasp-4b              & 	$1.33823187^{+0.00000025}_{-0.00000025}$ & 	$54823.591920^{+0.000028}_{-0.000028}$ & 	- & 	- & 	$88.80^{+0.61}_{-0.43}$ & 	$0.930^{+0.050}_{-0.050}$ & 	 \\ 
wasp-50b             & 	$1.9550959^{+0.0000051}_{-0.0000051}$ & 	$55558.61197^{+0.00021}_{-0.00015}$ & 	$0.0090^{+0.0060}_{-0.0060}$ & 	$44^{+80}_{-80}$ & 	$84.74^{+0.24}_{-0.24}$ & 	$0.861^{+0.057}_{-0.057}$ & 	 \\ 
wasp-52b             & 	$1.7497798^{+0.0000012}_{-0.0000012}$ & 	$55793.681430^{+0.000090}_{-0.000090}$ & 	- & 	- & 	$85.35^{+0.20}_{-0.20}$ & 	$0.870^{+0.030}_{-0.030}$ & 	 \\ 
wasp-5b              & 	$1.6284246^{+0.0000013}_{-0.0000013}$ & 	$54375.62494^{+0.00024}_{-0.00024}$ & 	- & 	- & 	$85.8^{+1.1}_{-1.1}$ & 	$1.000^{+0.060}_{-0.060}$ & 	 \\ 
wasp-65b             & 	$2.3114243^{+0.0000015}_{-0.0000015}$ & 	$56110.68772^{+0.00015}_{-0.00015}$ & 	- & 	- & 	$88.80^{+0.80}_{-0.70}$ & 	$0.93^{+0.12}_{-0.16}$ & 	 \\ 
wasp-68b             & 	$5.084298^{+0.000015}_{-0.000015}$ & 	$56064.86356^{+0.00060}_{-0.00060}$ & 	- & 	- & 	$88.1^{+1.3}_{-1.3}$ & 	$1.230^{+0.030}_{-0.030}$ & 	 \\ 
wasp-77ab            & 	$1.3600309^{+0.0000020}_{-0.0000020}$ & 	$55870.44977^{+0.00014}_{-0.00014}$ & 	- & 	- & 	$89.40^{+0.40}_{-0.70}$ & 	$1.002^{+0.045}_{-0.045}$ & 	 \\ 
wasp-80b             & 	$3.0678504^{+0.0000023}_{-0.0000027}$ & 	$56125.417510^{+0.000052}_{-0.000067}$ & 	- & 	- & 	$89.92^{+0.070}_{-0.12}$ & 	$0.580^{+0.050}_{-0.050}$ & 	 \\ 
wasp-98b             & 	$2.9626400^{+0.0000013}_{-0.0000013}$ & 	$56333.39130^{+0.00010}_{-0.00010}$ & 	- & 	- & 	$86.30^{+0.10}_{-0.10}$ & 	$0.690^{+0.060}_{-0.060}$ & 	 \\ 
wts-2b               & 	$1.0187074^{+0.0000070}_{-0.0000070}$ & 	$54317.81264^{+0.00070}_{-0.00070}$ & 	- & 	- & 	$83.43^{+0.53}_{-0.53}$ & 	$0.820^{+0.082}_{-0.082}$ & 	 \\

\hline
\hline
\end{tabular}
\end{center}
\label{tab:literature}
\end{table*}%

\begin{table*}[]
\setlength{\extrarowheight}{3pt}
\caption{Properties of the archival radial velocity data used in the 46 systems analyzed.}
\scriptsize
\begin{center}
\begin{tabular}{lccrrc}
\hline
\hline

System & $N_{\rm meas}$ & $N_{\rm inst}$ & Timespan (days) & RV scatter (m/s) & Ref. \\ 
\hline

corot-1b &          17 &             3 &         655.8 &          83.4 &        [0],[1]         \\ 
corot-2b &          35 &             5 &          83.1 &          46.4 &        [2],[3]         \\ 
corot-15b &         16 &             2 &          88.7 &         345.7 &        [4]         \\ 
corot-18b &         28 &             4 &         106.0 &          51.3 &        [5] 	    \\ 
epic211089792b &    42 &             5 &         113.7 &          17.1 &        [6],[7]         \\ 
gj3470b &           61 &             1 &        1153.9 &           4.4 &        [8] 	    \\ 
hat-p-12b &         23 &             1 &        2158.0 &           4.1 &        [9] 	    \\ 
hat-p-20b &         15 &             1 &        1704.3 &          11.2 &        [9] 	    \\ 
hat-p-23b &         27 &             2 &         796.5 &          27.3 &        [10],[11]       \\ 
hat-p-36b &         15 &             2 &         798.6 &          33.7 &        [12],[13]       \\ 
hat-p-54b &         17 &             2 &         410.0 &          51.7 &        [14]    \\ 
hats-2b &           18 &             3 &         334.2 &          99.3 &        [15]    \\ 
hats-9b &           21 &             3 &         258.3 &          22.9 &        [16]    \\ 
k2-30b &            22 &             4 &          61.0 &          16.2 &        [17],[18]       \\ 
kepler-17b &        20 &             2 &         273.7 &          50.1 &        [20],[21]       \\ 
ogle-tr-113b &       8 &             1 &           7.0 &          44.9 &        [22]    \\ 
ogle-tr-56b &       12 &             1 &          46.9 &          52.9 &        [23]    \\ 
qatar-1b &          20 &             2 &         663.9 &          31.2 &        [24],[25]       \\ 
qatar-2b &          44 &             1 &         153.7 &          82.0 &        [26]   \\ 
tres-3b &           13 &             2 &        1946.8 &          13.5 &        [9],[27]        \\ 
tres-5b &            8 &             1 &         217.1 &          19.1 &        [28]    \\ 
wasp-103b &         20 &             2 &         350.5 &          23.1 &        [29],[30]       \\ 
wasp-104b &         21 &             2 &         168.9 &          15.5 &        [31]    \\ 
wasp-12b &          47 &             2 &        2128.6 &          14.6 &        [33],[9]        \\ 
wasp-135b &         18 &             1 &         124.8 &          17.2 &        [34]    \\ 
wasp-13b &          11 &             1 &           4.2 &          11.3 &        [35]    \\ 
wasp-18b &          81 &             5 &        1849.2 &          12.7 &        [36],[37],[9],[38]      \\ 
wasp-19b &          44 &             2 &         722.2 &          25.1 &        [36],[39]       \\ 
wasp-23b &          59 &             2 &         584.7 &          17.6 &        [40]    \\ 
wasp-2b &           56 &             5 &        2198.2 &          20.8 &        [41],[9],[38]           \\ 
wasp-35b &          12 &             2 &          39.8 &          10.9 &        [42]    \\ 
wasp-36b &          19 &             1 &         306.1 &          17.2 &        [43]    \\ 
wasp-42b &          54 &             2 &         400.6 &           8.2 &        [44]    \\ 
wasp-43b &           8 &             1 &          37.9 &           6.7 &        [45]    \\ 
wasp-46b &          16 &             1 &         119.9 &          33.3 &        [46]    \\ 
wasp-49b &          24 &             1 &         775.0 &          13.6 &        [44]    \\ 
wasp-4b &           57 &             4 &        2172.4 &          17.2 &        [9],[40],[47]           \\ 
wasp-50b &          15 &             1 &          30.0 &          12.7 &        [45]    \\ 
wasp-52b &          56 &             5 &         444.8 &          15.2 &        [48]    \\ 
wasp-5b &           46 &             3 &         397.2 &          13.2 &        [49],[40]       \\ 
wasp-65b &          17 &             1 &         321.1 &          10.6 &        [50]    \\ 
wasp-68b &          43 &             1 &         823.8 &          12.7 &        [51]    \\ 
wasp-77ab &         18 &             2 &         848.9 &          10.6 &        [52]    \\ 
wasp-80b &          47 &             2 &         418.9 &          22.0 &        [53]    \\ 
wasp-98b &          14 &             1 &         364.9 &          25.0 &        [54]    \\ 
wts-2b &             6 &             1 &          53.9 &          24.9 &        [55]    \\ 

\hline
\hline
\end{tabular}
\tablefoot{[0] \cite{barge08}, [1] \cite{pont09}, [2] \cite{alonso08}, [3] \cite{bouchy08}, [4] \cite{bouchy10}, [5] \cite{hebrard11}, [6] \cite{johnson16}, [7] \cite{santerne16b}, [8] \cite{bonfils12}, [9] \cite{knutson14}, [10] \cite{bakos10}, [11] \cite{moutou11b}, [12] \cite{mancini15b}, [13] \cite{bakos12}, [14] \cite{bakos15}, [15] \cite{mohler-fischer13}, [16] \cite{brahm15}, [17] \cite{johnson16}, [18] \cite{lillo-box16b}, [19] \cite{batalha11}, [20] \cite{desert11b}, [21] \cite{bonomo12}, [22] \cite{pont05}, [23] \cite{bouchy05b}, [24] \cite{alsubai11}, [25] \cite{covino13}, [26] \cite{bryan11}, [27] \cite{odonovan07}, [28] \cite{mandushev11}, [29] \cite{addison16}, [30] \cite{gillon14}, [31] \cite{smith14}, [32] \cite{christian08}, [33] \cite{hebb09}, [34] \cite{spake16}, [35] \cite{skillen09}, [36] \cite{albrecht12}, [37] \cite{hellier09}, [38] \cite{triaud10}, [39] \cite{hebb10}, [40] \cite{triaud11}, [41] \cite{collier-cameron07}, [42] \cite{enoch11}, [43] \cite{smith12}, [44] \cite{lendl12}, [45] \cite{gillon12}, [46] \cite{anderson12}, [47] \cite{wilson08}, [48] \cite{hebrard13}, [49] \cite{anderson08}, [50] \cite{gomez-maqueo13}, [51] \cite{delrez14}, [52] \cite{maxted16}, [53] \cite{triaud13}, [54] \cite{hellier14}, [55] \cite{birkby14}.}
\end{center}
\label{tab:rvdata}
\end{table*}%

\begin{table*}[]
\setlength{\extrarowheight}{2pt}
\small
\caption{Adopted $e\cos{\omega}$ values to set the priors on the $c$ parameter for the 20 planets with detected secondary eclipse. }
\begin{center}
\begin{tabular}{lcl}
\hline
\hline
Planet & $e\cos{\omega}$ & Reference           \\ \hline
corot-1b     &  0.0020	  $\pm$0.0029	 & \cite{gillon09}		    \\
             &            $<$0.002	 & \cite{deming11}		    \\
             &  	  $<$0.014	 & \cite{alonso09}		    \\
             &  -0.0025	  $\pm$0.0010	 & $^{\dagger}$\cite{parviainen13}    \\
corot-15b    &  -0.00249  $\pm$0.0005	 & $^{\dagger}$\cite{parviainen13}    \\
corot-18b    &  -0.0154	  $\pm$0.001	 & $^{\dagger}$\cite{parviainen13}    \\
corot-2b     &  0.0005	  $\pm$0.0010	 & $^{\dagger}$\cite{parviainen13}    \\
             &  -0.0030	  $\pm$0.0004	 & \cite{deming11}		    \\
             &  -0.00291  $\pm$0.00063	 & \cite{gillon10}		    \\
             &  -0.0025	  $\pm$0.0015	 & \cite{snellen10}		    \\
hat-p-23b    &  -0.0011	  $\pm$0.0065	 & \cite{orourke14}		    \\
ogle-tr-56b  &  -0.00147  $\pm$0.0049	 & \cite{sing09}		    \\
tres-3b	     &  	  $<$0.0019	 & \cite{fressin10}		    \\
             &  0.0029	  $\pm$0.0022	 & \cite{croll10}		    \\
 	     &  -0.0066	  $\pm$0.0021	 & \cite{demooij09}		    \\
wasp-10b     &  -0.0552   $\pm$0.0007	 & \cite{kammer15}		    \\
             &  -0.0044   $\pm$0.0004	 & \cite{cruz14}		    \\
wasp-12b     &  0.000097  $\pm$0.000401 & $^{\dagger}$\cite{croll10}	    \\
             &  0.0005	  $\pm$0.0010	 & $^{\dagger}$\cite{fohring13}	    \\
             &  0.0014	  $\pm$0.0007	 & \cite{campo11}		    \\
             &  0.0050	  $\pm$0.0037	 & $^{\dagger}$\cite{lopez-morales10}  \\
wasp-18b     &  0.012	  $\pm$0.008	 & \cite{zhou15}		    \\
             &  -0.0003	  $\pm$0.0002	 & \cite{nymeyer11}		    \\
wasp-19b     &  -0.0056	  $\pm$0.0070	 & \cite{zhou14}		    \\
             &  -0.0049	  $\pm$0.0023	 & \cite{burton12}		    \\
wasp-2b	     &  -0.001	  $\pm$0.001	 & \cite{zhou15}		    \\
             &  -0.0013	  $\pm$0.0009	 & \cite{wheatley10}		    \\
wasp-36b     &  0.004	  $\pm$0.006	 & \cite{zhou15}		    \\
wasp-4b	     &  0.00030	  $\pm$0.00086	 & \cite{beerer11}		    \\
  	     &  -0.001	  $\pm$0.003	 & \cite{zhou15}		    \\
wasp-43b     &  -0.0062	  $\pm$0.0024	 & \cite{zhou14}		    \\
wasp-46b     &  0.004	  $\pm$0.002	 & \cite{zhou15}		    \\
wasp-5b	     &  0.008	  $\pm$0.002	 & \cite{zhou15}		    \\
 	     &  0.0025	  $\pm$0.0012	 & \cite{baskin13}		    \\
kepler-17b   &  0.0	  $<$0.011	 & \cite{desert11}		    \\
qatar-1b     &  -0.0123	  $\pm$0.0252	 & \cite{cruz16}		    \\
hat-p-20b    &  0.0132	  $\pm$0.0006	 &\cite{deming15}	    \\

\hline
\hline
\end{tabular}
\tablefoot{
Given the adopted reference frame for the radial velocity the priors for $c$ are set as $\mathcal{G}(-e\cos{\omega},5\sigma_{e\cos{\omega}})$.\\
$^{\dagger}$In these cases the value for $e\cos{\omega}$ was derived from the phase shift determined by the authors and the inclination value from exoplanet.org.}
\end{center}
\label{tab:ecosw}
\end{table*}%


\begin{table*}[]
\setlength{\extrarowheight}{3pt}
\caption{Derived parameters for the 46 planetary systems analyzed and the nine models tested. Hyperparameters of the Gaussian process kernel, radial velocity offsets, and jitter are not included here (see Table\ref{tab:hyperparams}). \label{tab:fiting_results}}
\scriptsize
\begin{center}
\begin{tabular}{llllllllll}
\hline
\hline

Object & $\gamma$ & $P$    & $T_0$          & $K$    & $\alpha$ & $c$ & $d$ \\ 
       & (km/s)       & (days) & (BJD-2400000) & $(m/s)$ &          &     &  \\ \hline

corot-1b             & 	$23.430^{+0.056}_{-0.060}$ & 	$1.508956^{+0.000019}_{-0.000018}$ & 	$54159.45319^{+0.00030}_{-0.00030}$ & 	$192^{+12}_{-12}$ & 	$-0.028^{+0.077}_{-0.073}$ & 	$-0.001^{+0.018}_{-0.018}$ & 	$0.051^{+0.065}_{-0.070}$ 	 \\ 
corot-2b             & 	$23.193^{+0.038}_{-0.039}$ & 	$1.7429961^{+0.0000052}_{-0.0000054}$ & 	$54706.4038^{+0.0089}_{-0.0098}$ & 	$582.9^{+2.3}_{-2.2}$ & 	$-0.021^{+0.060}_{-0.059}$ & 	$0.0019^{+0.0027}_{-0.0027}$ & 	$-0.032^{+0.042}_{-0.036}$ 	 \\ 
corot-15b            & 	$2.27^{+0.14}_{-0.13}$ & 	$3.060360^{+0.000093}_{-0.000088}$ & 	$54753.5606^{+0.0031}_{-0.0033}$ & 	$7440^{+140}_{-140}$ & 	$0.011^{+0.032}_{-0.034}$ & 	$0.0022^{+0.0026}_{-0.0025}$ & 	$-0.015^{+0.016}_{-0.015}$ 	 \\ 
corot-18b            & 	$29.535^{+0.027}_{-0.027}$ & 	$1.9000693^{+0.0000081}_{-0.0000082}$ & 	$55321.72417^{+0.00058}_{-0.00054}$ & 	$578^{+20}_{-20}$ & 	$0.039^{+0.045}_{-0.045}$ & 	$0.0154^{+0.0043}_{-0.0049}$ & 	$0.044^{+0.040}_{-0.045}$ 	 \\ 
epic211089792b       & 	$32.91^{+0.26}_{-0.28}$ & 	$3.2588322^{+0.0000059}_{-0.0000059}$ & 	$53219.0093^{+0.0065}_{-0.0063}$ & 	$97.0^{+4.7}_{-4.9}$ & 	$-0.15^{+0.11}_{-0.10}$ & 	$0.000^{+0.060}_{-0.066}$ & 	$0.036^{+0.051}_{-0.051}$ 	 \\ 
gj3470b              & 	$26.5156^{+0.0052}_{-0.0054}$ & 	$3.33674^{+0.00014}_{-0.00014}$ & 	$56090.47688^{+0.00042}_{-0.00044}$ & 	$7.96^{+0.93}_{-0.88}$ & 	$-0.61^{+0.26}_{-0.23}$ & 	$-0.19^{+0.11}_{-0.091}$ & 	$0.06^{+0.12}_{-0.11}$ 	 \\ 
hat-p-12b            & 	$0.0005^{+0.0043}_{-0.0042}$ & 	$3.2130601^{+0.0000063}_{-0.0000060}$ & 	$54419.19561^{+0.00058}_{-0.00061}$ & 	$37.0^{+2.9}_{-3.3}$ & 	$-0.18^{+0.11}_{-0.14}$ & 	$-0.086^{+0.058}_{-0.062}$ & 	$-0.006^{+0.053}_{-0.049}$ 	 \\ 
hat-p-20b            & 	$0.069^{+0.015}_{-0.014}$ & 	$2.875322^{+0.0000099}_{-0.000010}$ & 	$56708.35628^{+0.00026}_{-0.00025}$ & 	$1245.2^{+6.2}_{-6.2}$ & 	$0.000^{+0.013}_{-0.013}$ & 	$-0.0147^{+0.0027}_{-0.0028}$ & 	$0.0088^{+0.0050}_{-0.0054}$ 	 \\ 
hat-p-23b            & 	$-0.006^{+0.019}_{-0.018}$ & 	$1.212863^{+0.000041}_{-0.000037}$ & 	$54852.26465^{+0.00054}_{-0.00056}$ & 	$358^{+15}_{-14}$ & 	$-0.120^{+0.094}_{-0.096}$ & 	$0.001^{+0.028}_{-0.026}$ & 	$-0.095^{+0.033}_{-0.038}$ 	 \\ 
hat-p-36b            & 	$0.013^{+0.034}_{-0.031}$ & 	$1.3273467^{+0.0000087}_{-0.0000086}$ & 	$55565.18142^{+0.00058}_{-0.00062}$ & 	$329^{+14}_{-18}$ & 	$0.25^{+0.24}_{-0.22}$ & 	$0.11^{+0.15}_{-0.13}$ & 	$-0.078^{+0.035}_{-0.044}$ 	 \\ 
hat-p-54b            & 	$-0.009^{+0.031}_{-0.033}$ & 	$3.799845^{+0.000042}_{-0.000041}$ & 	$56299.30369^{+0.00072}_{-0.00072}$ & 	$130^{+15}_{-15}$ & 	$0.14^{+0.59}_{-0.72}$ & 	$-0.00^{+0.21}_{-0.23}$ & 	$-0.16^{+0.26}_{-0.25}$ 	 \\ 
hats-2b              & 	$-0.010^{+0.081}_{-0.069}$ & 	$1.3541330^{+0.0000029}_{-0.0000029}$ & 	$55954.58575^{+0.00028}_{-0.00027}$ & 	$229^{+35}_{-43}$ & 	$0.05^{+0.46}_{-0.49}$ & 	$-0.03^{+0.20}_{-0.21}$ & 	$0.11^{+0.16}_{-0.16}$ 	 \\ 
hats-9b              & 	$-0.013^{+0.028}_{-0.029}$ & 	$1.915308^{+0.000015}_{-0.000016}$ & 	$56124.2591^{+0.0027}_{-0.0028}$ & 	$131^{+11}_{-11}$ & 	$-0.26^{+0.46}_{-0.42}$ & 	$-0.12^{+0.17}_{-0.15}$ & 	$0.02^{+0.18}_{-0.16}$ 	 \\ 
k2-30b               & 	$35.422^{+0.037}_{-0.032}$ & 	$4.098510^{+0.000088}_{-0.000082}$ & 	$57063.80706^{+0.00088}_{-0.00087}$ & 	$79.6^{+7.1}_{-7.5}$ & 	$0.02^{+0.25}_{-0.28}$ & 	$0.03^{+0.11}_{-0.11}$ & 	$0.003^{+0.088}_{-0.080}$ 	 \\ 
kepler-17b           & 	$-0.017^{+0.029}_{-0.028}$ & 	$1.48571080^{+0.00000059}_{-0.00000058}$ & 	$55185.678036^{+0.000074}_{-0.000081}$ & 	$397^{+18}_{-18}$ & 	$0.033^{+0.074}_{-0.072}$ & 	$0.015^{+0.029}_{-0.031}$ & 	$0.023^{+0.057}_{-0.054}$ 	 \\ 
ogle-tr-113b         & 	$-7.989^{+0.050}_{-0.049}$ & 	$1.4324769^{+0.0000037}_{-0.0000036}$ & 	$53471.7782^{+0.0015}_{-0.0015}$ & 	$274^{+44}_{-61}$ & 	$-0.13^{+0.26}_{-0.29}$ & 	$-0.02^{+0.13}_{-0.13}$ & 	$-0.11^{+0.19}_{-0.14}$ 	 \\ 
ogle-tr-56b          & 	$-48.320^{+0.081}_{-0.083}$ & 	$1.2119087^{+0.0000029}_{-0.0000029}$ & 	$53936.5979^{+0.0029}_{-0.0030}$ & 	$177^{+36}_{-37}$ & 	$-0.19^{+0.22}_{-0.26}$ & 	$0.000^{+0.024}_{-0.024}$ & 	$-0.10^{+0.22}_{-0.30}$ 	 \\ 
qatar-1b             & 	$0.126^{+0.053}_{-0.051}$ & 	$1.4200246^{+0.0000022}_{-0.0000021}$ & 	$56157.42202^{+0.00031}_{-0.00030}$ & 	$260.2^{+6.6}_{-7.2}$ & 	$-0.008^{+0.094}_{-0.076}$ & 	$0.001^{+0.039}_{-0.032}$ & 	$-0.003^{+0.025}_{-0.022}$ 	 \\ 
qatar-2b             & 	$-0.210^{+0.028}_{-0.029}$ & 	$1.337119^{+0.000011}_{-0.000011}$ & 	$55624.26680^{+0.00033}_{-0.00034}$ & 	$561.2^{+7.8}_{-7.6}$ & 	$-0.028^{+0.038}_{-0.038}$ & 	$-0.010^{+0.017}_{-0.018}$ & 	$0.009^{+0.012}_{-0.012}$ 	 \\ 
tres-3b              & 	$-0.022^{+0.026}_{-0.026}$ & 	$1.3061860^{+0.0000011}_{-0.0000011}$ & 	$54538.58064^{+0.00062}_{-0.00061}$ & 	$340^{+12}_{-13}$ & 	$-0.066^{+0.036}_{-0.035}$ & 	$0.0006^{+0.0061}_{-0.0058}$ & 	$0.108^{+0.035}_{-0.033}$ 	 \\ 
tres-5b              & 	$-0.211^{+0.029}_{-0.029}$ & 	$1.4822444^{+0.0000022}_{-0.0000021}$ & 	$55443.25151^{+0.00032}_{-0.00033}$ & 	$342^{+23}_{-28}$ & 	$-0.03^{+0.53}_{-0.52}$ & 	$-0.03^{+0.18}_{-0.18}$ & 	$0.00^{+0.091}_{-0.10}$ 	 \\ 
wasp-103b            & 	$-42.017^{+0.020}_{-0.020}$ & 	$0.925545^{+0.000058}_{-0.000059}$ & 	$56459.5996^{+0.0022}_{-0.0025}$ & 	$251^{+18}_{-19}$ & 	$-0.12^{+0.29}_{-0.28}$ & 	$-0.05^{+0.095}_{-0.10}$ & 	$-0.036^{+0.086}_{-0.086}$ 	 \\ 
wasp-104b            & 	$28.8303^{+0.0078}_{-0.0075}$ & 	$1.7554141^{+0.0000078}_{-0.0000082}$ & 	$56406.11127^{+0.00035}_{-0.00035}$ & 	$202.2^{+7.7}_{-8.0}$ & 	$0.004^{+0.070}_{-0.070}$ & 	$0.001^{+0.029}_{-0.029}$ & 	$0.015^{+0.033}_{-0.036}$ 	 \\ 
wasp-12b             & 	$19.086^{+0.011}_{-0.0099}$ & 	$1.0914170^{+0.0000032}_{-0.0000031}$ & 	$54508.97608^{+0.00076}_{-0.00079}$ & 	$218.6^{+3.6}_{-3.5}$ & 	$0.015^{+0.026}_{-0.028}$ & 	$0.0001^{+0.0049}_{-0.0046}$ & 	$0.0064^{+0.0056}_{-0.0058}$ 	 \\ 
wasp-135b            & 	$-32.35^{+0.12}_{-0.095}$ & 	$1.4013795^{+0.0000023}_{-0.0000025}$ & 	$55230.9904^{+0.0026}_{-0.0027}$ & 	$347.6^{+7.1}_{-7.1}$ & 	$-0.002^{+0.060}_{-0.059}$ & 	$-0.002^{+0.024}_{-0.024}$ & 	$0.050^{+0.022}_{-0.022}$ 	 \\ 
wasp-13b             & 	$9.839^{+0.079}_{-0.082}$ & 	$4.353010^{+0.000039}_{-0.000039}$ & 	$55575.5131^{+0.0047}_{-0.0045}$ & 	$43^{+11}_{-21}$ & 	$-0.24^{+0.66}_{-0.48}$ & 	$-0.22^{+0.31}_{-0.23}$ & 	$0.20^{+0.27}_{-0.24}$ 	 \\ 
wasp-18b             & 	$3.2004^{+0.0097}_{-0.0094}$ & 	$0.94145224^{+0.00000061}_{-0.00000061}$ & 	$54221.4816^{+0.0012}_{-0.0012}$ & 	$1814.0^{+3.8}_{-3.8}$ & 	$-0.0022^{+0.0083}_{-0.0090}$ & 	$-0.0003^{+0.0021}_{-0.0020}$ & 	$0.0101^{+0.0014}_{-0.0015}$ 	 \\ 
wasp-19b             & 	$20.804^{+0.014}_{-0.020}$ & 	$0.78883992^{+0.00000093}_{-0.00000088}$ & 	$55168.96801^{+0.00027}_{-0.00027}$ & 	$254.7^{+6.1}_{-6.4}$ & 	$0.034^{+0.033}_{-0.033}$ & 	$0.006^{+0.010}_{-0.011}$ & 	$0.039^{+0.023}_{-0.024}$ 	 \\ 
wasp-23b             & 	$5.662^{+0.0098}_{-0.011}$ & 	$2.9444257^{+0.0000037}_{-0.0000037}$ & 	$55320.12363^{+0.00038}_{-0.00039}$ & 	$146.0^{+2.7}_{-2.7}$ & 	$-0.001^{+0.054}_{-0.051}$ & 	$-0.003^{+0.023}_{-0.022}$ & 	$0.032^{+0.018}_{-0.018}$ 	 \\ 
wasp-2b              & 	$-27.861^{+0.012}_{-0.011}$ & 	$2.1522213^{+0.0000013}_{-0.0000013}$ & 	$53991.51529^{+0.00050}_{-0.00052}$ & 	$149.9^{+4.5}_{-4.3}$ & 	$-0.011^{+0.025}_{-0.018}$ & 	$0.0016^{+0.0032}_{-0.0036}$ & 	$0.027^{+0.028}_{-0.030}$ 	 \\ 
wasp-35b             & 	$17.733^{+0.026}_{-0.024}$ & 	$3.1615750^{+0.0000063}_{-0.0000061}$ & 	$55531.47905^{+0.00043}_{-0.00044}$ & 	$88^{+11}_{-16}$ & 	$0.08^{+0.21}_{-0.20}$ & 	$0.02^{+0.12}_{-0.14}$ & 	$0.09^{+0.20}_{-0.15}$ 	 \\ 
wasp-36b             & 	$-13.211^{+0.012}_{-0.012}$ & 	$1.5373677^{+0.0000079}_{-0.0000080}$ & 	$55569.83734^{+0.00027}_{-0.00029}$ & 	$387.2^{+7.5}_{-7.2}$ & 	$0.092^{+0.042}_{-0.043}$ & 	$0.022^{+0.020}_{-0.020}$ & 	$0.007^{+0.022}_{-0.023}$ 	 \\ 
wasp-42b             & 	$-4.7726^{+0.0095}_{-0.0094}$ & 	$4.981688^{+0.000023}_{-0.000023}$ & 	$55650.56716^{+0.00066}_{-0.00069}$ & 	$63.6^{+2.0}_{-1.9}$ & 	$0.015^{+0.057}_{-0.057}$ & 	$0.063^{+0.024}_{-0.023}$ & 	$-0.025^{+0.037}_{-0.036}$ 	 \\ 
wasp-43b             & 	$-3.591^{+0.016}_{-0.015}$ & 	$0.8134768^{+0.0000018}_{-0.0000019}$ & 	$55726.54335^{+0.00034}_{-0.00035}$ & 	$556^{+10}_{-9.0}$ & 	$-0.000^{+0.026}_{-0.029}$ & 	$0.0025^{+0.0091}_{-0.0094}$ & 	$0.012^{+0.011}_{-0.010}$ 	 \\ 
wasp-46b             & 	$-3.779^{+0.013}_{-0.013}$ & 	$1.4303697^{+0.0000069}_{-0.0000069}$ & 	$55392.31553^{+0.00059}_{-0.00057}$ & 	$383^{+14}_{-13}$ & 	$-0.012^{+0.053}_{-0.055}$ & 	$-0.005^{+0.010}_{-0.010}$ & 	$-0.028^{+0.037}_{-0.037}$ 	 \\ 
wasp-49b             & 	$41.6890^{+0.0061}_{-0.0061}$ & 	$2.781739^{+0.000017}_{-0.000017}$ & 	$55580.59437^{+0.00087}_{-0.00087}$ & 	$56.1^{+4.0}_{-4.5}$ & 	$-0.03^{+0.17}_{-0.18}$ & 	$-0.005^{+0.074}_{-0.073}$ & 	$-0.017^{+0.089}_{-0.077}$ 	 \\ 
wasp-4b              & 	$57.732^{+0.013}_{-0.011}$ & 	$1.33823185^{+0.00000075}_{-0.00000075}$ & 	$54823.591921^{+0.000084}_{-0.000079}$ & 	$240.9^{+4.3}_{-4.5}$ & 	$-0.021^{+0.019}_{-0.020}$ & 	$-0.0012^{+0.0073}_{-0.0068}$ & 	$-0.000^{+0.013}_{-0.012}$ 	 \\ 
wasp-50b             & 	$25.526^{+0.011}_{-0.013}$ & 	$1.955096^{+0.000015}_{-0.000015}$ & 	$55558.61193^{+0.00055}_{-0.00053}$ & 	$255.3^{+5.8}_{-6.4}$ & 	$0.02^{+0.096}_{-0.10}$ & 	$-0.002^{+0.044}_{-0.046}$ & 	$-0.005^{+0.053}_{-0.065}$ 	 \\ 
wasp-52b             & 	$-0.8871^{+0.0071}_{-0.0073}$ & 	$1.7497796^{+0.0000039}_{-0.0000039}$ & 	$55793.68142^{+0.00028}_{-0.00028}$ & 	$82.2^{+3.5}_{-1.5}$ & 	$0.01^{+0.11}_{-0.11}$ & 	$-0.025^{+0.049}_{-0.045}$ & 	$-0.048^{+0.046}_{-0.046}$ 	 \\ 
wasp-5b              & 	$20.0128^{+0.0084}_{-0.0090}$ & 	$1.6284243^{+0.0000040}_{-0.0000039}$ & 	$54375.62494^{+0.00071}_{-0.00071}$ & 	$267.1^{+1.4}_{-1.5}$ & 	$0.005^{+0.012}_{-0.013}$ & 	$-0.0026^{+0.0045}_{-0.0045}$ & 	$-0.0019^{+0.0057}_{-0.0067}$ 	 \\ 
wasp-65b             & 	$-3.187^{+0.014}_{-0.014}$ & 	$2.3114244^{+0.0000043}_{-0.0000043}$ & 	$56110.68770^{+0.00047}_{-0.00043}$ & 	$248.4^{+7.2}_{-7.1}$ & 	$-0.012^{+0.077}_{-0.075}$ & 	$0.001^{+0.037}_{-0.038}$ & 	$0.024^{+0.026}_{-0.026}$ 	 \\ 
wasp-68b             & 	$28.3097^{+0.0022}_{-0.0022}$ & 	$5.084304^{+0.000045}_{-0.000047}$ & 	$56064.8637^{+0.0017}_{-0.0018}$ & 	$97.5^{+2.7}_{-2.6}$ & 	$0.040^{+0.066}_{-0.060}$ & 	$0.003^{+0.026}_{-0.025}$ & 	$0.016^{+0.026}_{-0.026}$ 	 \\ 
wasp-77ab            & 	$1.663^{+0.013}_{-0.012}$ & 	$1.3600310^{+0.0000056}_{-0.0000060}$ & 	$55870.44980^{+0.00042}_{-0.00043}$ & 	$325.1^{+4.2}_{-4.1}$ & 	$0.045^{+0.040}_{-0.037}$ & 	$0.016^{+0.012}_{-0.012}$ & 	$0.023^{+0.019}_{-0.017}$ 	 \\ 
wasp-80b             & 	$10.2355^{+0.0068}_{-0.0067}$ & 	$3.0678502^{+0.0000076}_{-0.0000070}$ & 	$56125.41751^{+0.00019}_{-0.00019}$ & 	$108.0^{+4.6}_{-4.7}$ & 	$0.038^{+0.085}_{-0.083}$ & 	$0.030^{+0.045}_{-0.046}$ & 	$0.004^{+0.036}_{-0.039}$ 	 \\ 
wasp-98b             & 	$-38.290^{+0.019}_{-0.018}$ & 	$2.9626401^{+0.0000041}_{-0.0000038}$ & 	$56333.39130^{+0.00032}_{-0.00030}$ & 	$148^{+15}_{-15}$ & 	$0.14^{+0.46}_{-0.48}$ & 	$0.09^{+0.15}_{-0.17}$ & 	$-0.04^{+0.11}_{-0.11}$ 	 \\ 
wts-2b               & 	$-20.06^{+0.12}_{-0.11}$ & 	$1.018704^{+0.000023}_{-0.000024}$ & 	$54317.8127^{+0.0021}_{-0.0021}$ & 	$191^{+62}_{-76}$ & 	$0.22^{+0.56}_{-0.72}$ & 	$-0.23^{+0.29}_{-0.35}$ & 	$0.13^{+0.30}_{-0.30}$ 	 \\

\hline
\hline
\end{tabular}
\end{center}
\end{table*}%


\begin{landscape}
\begin{table}[]
\setlength{\extrarowheight}{1pt}
\caption{RResults for the Gaussian process hyperparameters, radial velocity offsets, and instrumental jitters for each of the 46 systems. \label{tab:hyperparams}}
\scriptsize
\begin{center}
\begin{tabular}{llllllll}
\hline
\hline

Object & $\eta_1$ & $\eta_2$    & $\eta_3$          & $\eta_4$    & slope & Instrumental offsets ($\phi$) and jitters ($\sigma$)  \\ 
       & (m/s)       & (days)   & (days)            &            & m/s/yr &  (m/s)   \\ \hline

corot-1b             & 	$8.7^{+6.3}_{-5.9}$ & 	$70^{+21}_{-29}$ & 	$63^{+26}_{-36}$ & 	$-1.9^{+1.6}_{-1.6}$ & 	$2^{+25}_{-26}$ & 	$\phi_{\rm H} =$$-93.64^{+103.81}_{-107.77}$,  	$\phi_{\rm HI} =$$25976.81^{+117.43}_{-130.26}$,  	$\sigma_{\rm S} =$$15^{+23}_{-11}$,  	$\sigma_{\rm H} =$$13^{+17}_{-7.5}$,  	$\sigma_{\rm HI} =$$54^{+30}_{-25}$ 	 \\ 
corot-2b             & 	$3.4^{+2.6}_{-2.6}$ & 	$80^{+9.5}_{-11}$ & 	$77^{+11}_{-11}$ & 	$3.51^{+0.54}_{-0.62}$ & 	$0^{+26}_{-26}$ & 	$\phi_{\rm S} =$$2^{+48}_{-48}$,  	$\phi_{\rm C} =$$106^{+64}_{-67}$,  	$\phi_{\rm H} =$$-50^{+46}_{-45}$,  	$\phi_{\rm H} =$$-49^{+41}_{-44}$,  	$\sigma_{\rm S} =$$32^{+33}_{-24}$,  	$\sigma_{\rm S} =$$34^{+42}_{-25}$,  	$\sigma_{\rm C} =$$55^{+31}_{-38}$,  	$\sigma_{\rm H} =$$12^{+14}_{-8.6}$,  	$\sigma_{\rm H} =$$14^{+17}_{-10}$ 	 \\ 
corot-15b            & 	$2.2^{+4.6}_{-4.4}$ & 	$50^{+32}_{-31}$ & 	$49^{+32}_{-31}$ & 	$0.2^{+2.8}_{-3.0}$ & 	$-4^{+26}_{-23}$ & 	$\phi_{\rm HI} =$$1118.90^{+176.56}_{-159.93}$,  	$\sigma_{\rm H} =$$49^{+36}_{-33}$,  	$\sigma_{\rm HI} =$$45^{+37}_{-30}$ 	 \\ 
corot-18b            & 	$0.46^{+0.64}_{-0.49}$ & 	$70^{+22}_{-39}$ & 	$56^{+27}_{-31}$ & 	$-0.1^{+2.8}_{-2.7}$ & 	$-4^{+27}_{-24}$ & 	$\phi_{\rm H} =$$-2^{+37}_{-41}$,  	$\phi_{\rm FI} =$$-144^{+53}_{-52}$,  	$\phi_{\rm H} =$$-6^{+51}_{-56}$,  	$\sigma_{\rm S} =$$26^{+28}_{-19}$,  	$\sigma_{\rm H} =$$36^{+34}_{-25}$,  	$\sigma_{\rm FI} =$$39^{+34}_{-27}$,  	$\sigma_{\rm H} =$$46^{+27}_{-25}$ 	 \\ 
epic211089792b       & 	$0.142^{+0.067}_{-0.037}$ & 	$55^{+31}_{-32}$ & 	$54^{+30}_{-33}$ & 	$-0.0^{+2.8}_{-2.9}$ & 	$-2^{+25}_{-23}$ & 	$\phi_{\rm FI} =$$164^{+14}_{-14}$,  	$\phi_{\rm HN} =$$-57^{+12}_{-12}$,  	$\sigma_{\rm S} =$$5.8^{+5.3}_{-4.1}$,  	$\sigma_{\rm FI} =$$17^{+16}_{-11}$,  	$\sigma_{\rm HN} =$$7.9^{+8.8}_{-5.6}$ 	 \\ 
gj3470b              & 	$0.109^{+0.015}_{-0.0071}$ & 	$88^{+9.1}_{-16}$ & 	$59^{+29}_{-34}$ & 	$-2.5^{+1.4}_{-1.2}$ & 	$-0.6^{+3.0}_{-3.1}$ & 	$\sigma_{\rm H} =$$1.03^{+0.98}_{-0.72}$ 	 \\ 
hat-p-12b            & 	$0.120^{+0.036}_{-0.016}$ & 	$58^{+29}_{-32}$ & 	$56^{+30}_{-34}$ & 	$-1.5^{+2.4}_{-1.9}$ & 	$-0.1^{+1.7}_{-1.7}$ & 	$\sigma_{\rm HI} =$$3.0^{+1.7}_{-1.5}$ 	 \\ 
hat-p-20b            & 	$0.17^{+0.11}_{-0.067}$ & 	$54^{+33}_{-33}$ & 	$50^{+32}_{-32}$ & 	$1.8^{+1.8}_{-4.0}$ & 	$-5.1^{+4.0}_{-3.8}$ & 	$\sigma_{\rm HI} =$$9.7^{+6.5}_{-6.3}$ 	 \\ 
hat-p-23b            & 	$0.42^{+0.48}_{-0.40}$ & 	$48^{+34}_{-30}$ & 	$54^{+32}_{-33}$ & 	$-0.2^{+2.8}_{-2.8}$ & 	$7^{+20}_{-22}$ & 	$\phi_{\rm S} =$$14226^{+68}_{-72}$,  	$\sigma_{\rm HI} =$$17^{+19}_{-12}$,  	$\sigma_{\rm S} =$$12^{+12}_{-8.5}$ 	 \\ 
hat-p-36b            & 	$2.1^{+2.7}_{-3.4}$ & 	$52^{+31}_{-30}$ & 	$51^{+35}_{-31}$ & 	$-1.2^{+2.6}_{-2.1}$ & 	$-3^{+27}_{-23}$ & 	$\phi_{\rm HN} =$$16373^{+87}_{-86}$,  	$\sigma_{\rm T} =$$23^{+22}_{-16}$,  	$\sigma_{\rm HN} =$$6^{+15}_{-4.7}$ 	 \\ 
hat-p-54b            & 	$1.6^{+1.3}_{-2.1}$ & 	$50^{+33}_{-30}$ & 	$57^{+30}_{-35}$ & 	$-1.4^{+2.3}_{-1.9}$ & 	$5^{+22}_{-26}$ & 	$\phi_{\rm HI} =$$-23^{+54}_{-53}$,  	$\sigma_{\rm T} =$$31^{+26}_{-20}$,  	$\sigma_{\rm HI} =$$39^{+38}_{-27}$ 	 \\ 
hats-2b              & 	$3.8^{+4.7}_{-8.0}$ & 	$48^{+35}_{-31}$ & 	$48^{+35}_{-30}$ & 	$1.0^{+2.3}_{-3.3}$ & 	$-9^{+30}_{-20}$ & 	$\phi_{\rm F} =$$-15.25^{+102.79}_{-85}$,  	$\phi_{\rm CY} =$$-9.73^{+111.16}_{-96}$,  	$\sigma_{\rm C} =$$59^{+26}_{-29}$,  	$\sigma_{\rm F} =$$57^{+30}_{-38}$,  	$\sigma_{\rm CY} =$$68^{+24}_{-37}$ 	 \\ 
hats-9b              & 	$0.49^{+0.62}_{-0.52}$ & 	$60^{+28}_{-34}$ & 	$52^{+33}_{-32}$ & 	$-1.0^{+2.2}_{-2.1}$ & 	$-2^{+26}_{-23}$ & 	$\phi_{\rm SU} =$$0^{+36}_{-39}$,  	$\phi_{\rm C} =$$-11^{+61}_{-67}$,  	$\sigma_{\rm F} =$$20^{+25}_{-14}$,  	$\sigma_{\rm SU} =$$7.2^{+7.2}_{-4.7}$,  	$\sigma_{\rm C} =$$43^{+34}_{-28}$ 	 \\ 
k2-30b               & 	$0.46^{+0.66}_{-0.48}$ & 	$88^{+9.2}_{-19}$ & 	$68^{+16}_{-37}$ & 	$-1.2^{+1.9}_{-2.0}$ & 	$3^{+22}_{-26}$ & 	$\phi_{\rm HN} =$$-199^{+35}_{-35}$,  	$\phi_{\rm H} =$$-201^{+22}_{-21}$,  	$\phi_{\rm S} =$$-88^{+32}_{-34}$,  	$\sigma_{\rm FI} =$$25^{+35}_{-19}$,  	$\sigma_{\rm HN} =$$22^{+32}_{-16}$,  	$\sigma_{\rm H} =$$5.4^{+9.3}_{-3.8}$,  	$\sigma_{\rm S} =$$12^{+16}_{-9.0}$ 	 \\ 
kepler-17b           & 	$0.31^{+0.33}_{-0.25}$ & 	$57^{+30}_{-34}$ & 	$53^{+32}_{-33}$ & 	$-0.0^{+2.9}_{-2.9}$ & 	$-1^{+25}_{-24}$ & 	$\phi_{\rm C} =$$24798^{+30}_{-30}$,  	$\sigma_{\rm HRS} =$$23^{+25}_{-16}$,  	$\sigma_{\rm C} =$$19^{+21}_{-13}$ 	 \\ 
ogle-tr-113b         & 	$1.3^{+2.5}_{-2.2}$ & 	$53^{+32}_{-33}$ & 	$49^{+33}_{-31}$ & 	$-0.2^{+3.2}_{-2.8}$ & 	$-1^{+26}_{-24}$ & 	$\sigma_{\rm FL} =$$56^{+29}_{-31}$ 	 \\ 
ogle-tr-56b          & 	$1.1^{+1.8}_{-1.7}$ & 	$59^{+29}_{-34}$ & 	$52^{+31}_{-33}$ & 	$-0.4^{+3.0}_{-2.6}$ & 	$2^{+23}_{-26}$ & 	$\sigma_{\rm FL} =$$33^{+32}_{-23}$ 	 \\ 
qatar-1b             & 	$0.55^{+0.82}_{-0.64}$ & 	$53^{+32}_{-32}$ & 	$53^{+32}_{-31}$ & 	$-0.7^{+2.7}_{-2.5}$ & 	$-2^{+25}_{-24}$ & 	$\phi_{\rm HN} =$$38174^{+60}_{-59}$,  	$\sigma_{\rm T} =$$42^{+23}_{-18}$,  	$\sigma_{\rm HN} =$$4.1^{+4.7}_{-2.9}$ 	 \\ 
qatar-2b             & 	$5.6^{+2.6}_{-2.2}$ & 	$86^{+10}_{-19}$ & 	$55^{+31}_{-33}$ & 	$-2.4^{+1.3}_{-1.3}$ & 	$2^{+24}_{-26}$ & 	$\sigma_{\rm T} =$$17.4^{+8.3}_{-8.4}$ 	 \\ 
tres-3b              & 	$0.50^{+0.58}_{-0.52}$ & 	$55^{+31}_{-33}$ & 	$54^{+31}_{-32}$ & 	$-0.6^{+2.9}_{-2.5}$ & 	$6.6^{+9.0}_{-7.9}$ & 	$\phi_{\rm HI} =$$-47^{+51}_{-45}$,  	$\sigma_{\rm HI} =$$13^{+13}_{-7.5}$,  	$\sigma_{\rm HI} =$$17^{+22}_{-12}$ 	 \\ 
tres-5b              & 	$0.58^{+0.91}_{-0.70}$ & 	$54^{+30}_{-31}$ & 	$48^{+32}_{-29}$ & 	$-0.5^{+3.2}_{-2.5}$ & 	$0^{+24}_{-24}$ & 	$\sigma_{\rm T} =$$23^{+31}_{-17}$ 	 \\ 
wasp-103b            & 	$0.34^{+0.37}_{-0.28}$ & 	$52^{+33}_{-32}$ & 	$56^{+31}_{-34}$ & 	$-0.3^{+2.8}_{-2.7}$ & 	$-0^{+24}_{-25}$ & 	$\phi_{\rm U} =$$637^{+70}_{-57}$,  	$\sigma_{\rm C} =$$15^{+16}_{-10}$,  	$\sigma_{\rm U} =$$18^{+24}_{-12}$ 	 \\ 
wasp-104b            & 	$0.18^{+0.11}_{-0.071}$ & 	$50^{+32}_{-32}$ & 	$53^{+33}_{-32}$ & 	$-0.7^{+2.4}_{-2.4}$ & 	$-2^{+26}_{-23}$ & 	$\phi_{\rm C} =$$281^{+12}_{-11}$,  	$\sigma_{\rm S} =$$5.4^{+5.9}_{-3.8}$,  	$\sigma_{\rm C} =$$17^{+11}_{-8.5}$ 	 \\ 
wasp-12b             & 	$0.191^{+0.081}_{-0.075}$ & 	$57^{+30}_{-33}$ & 	$50^{+33}_{-31}$ & 	$-0.3^{+3.3}_{-2.8}$ & 	$-3.0^{+3.8}_{-3.7}$ & 	$\phi_{\rm HI} =$$19061^{+16}_{-15}$,  	$\sigma_{\rm S} =$$8.1^{+5.0}_{-4.8}$,  	$\sigma_{\rm HI} =$$7.3^{+3.7}_{-2.9}$ 	 \\ 
wasp-135b            & 	$0.20^{+0.13}_{-0.092}$ & 	$52^{+31}_{-33}$ & 	$47^{+34}_{-30}$ & 	$1.7^{+1.8}_{-3.9}$ & 	$5^{+21}_{-26}$ & 	$\sigma_{\rm S} =$$11.6^{+7.8}_{-7.4}$ 	 \\ 
wasp-13b             & 	$0.7^{+1.5}_{-0.98}$ & 	$52^{+33}_{-33}$ & 	$55^{+30}_{-33}$ & 	$-0.2^{+2.8}_{-2.7}$ & 	$-0^{+26}_{-25}$ & 	$\sigma_{\rm S} =$$14.6^{+9.7}_{-6.0}$ 	 \\ 
wasp-18b             & 	$0.132^{+0.041}_{-0.026}$ & 	$36^{+37}_{-20}$ & 	$42^{+22}_{-19}$ & 	$3.25^{+0.66}_{-0.87}$ & 	$-1.9^{+5.7}_{-5.9}$ & 	$\phi_{\rm C} =$$-132.0^{+5.8}_{-5.2}$,  	$\phi_{\rm H} =$$-143.9^{+8.7}_{-8.6}$,  	$\phi_{\rm HI} =$$2639^{+21}_{-21}$,  	$\phi_{\rm PFS} =$$3025^{+25}_{-24}$,  	$\sigma_{\rm C} =$$4.9^{+6.0}_{-3.4}$,  	$\sigma_{\rm C} =$$3.7^{+3.8}_{-2.5}$,  	$\sigma_{\rm H} =$$3.5^{+3.4}_{-2.4}$,  	$\sigma_{\rm HI} =$$9^{+12}_{-6.1}$,  	$\sigma_{\rm PFS} =$$9.5^{+2.6}_{-2.3}$ 	 \\ 
wasp-19b             & 	$0.20^{+0.13}_{-0.094}$ & 	$50^{+34}_{-31}$ & 	$45^{+36}_{-31}$ & 	$1.7^{+2.2}_{-4.0}$ & 	$19^{+12}_{-18}$ & 	$\phi_{\rm PFS} =$$854^{+27}_{-31}$,  	$\sigma_{\rm C} =$$8.9^{+7.3}_{-6.1}$,  	$\sigma_{\rm PFS} =$$25^{+11}_{-10}$ 	 \\ 
wasp-23b             & 	$0.155^{+0.079}_{-0.048}$ & 	$64^{+25}_{-32}$ & 	$58^{+28}_{-34}$ & 	$-1.9^{+1.8}_{-1.6}$ & 	$-14.4^{+9.8}_{-9.6}$ & 	$\phi_{\rm H} =$$-26.5^{+9.5}_{-9.4}$,  	$\sigma_{\rm C} =$$3.7^{+3.9}_{-2.5}$,  	$\sigma_{\rm H} =$$3.0^{+2.1}_{-1.9}$ 	 \\ 
wasp-2b              & 	$0.155^{+0.093}_{-0.051}$ & 	$63^{+8.1}_{-14}$ & 	$71^{+10}_{-27}$ & 	$1.9^{+1.5}_{-1.7}$ & 	$-3^{+12}_{-12}$ & 	$\phi_{\rm S} =$$0^{+9.6}_{-10}$,  	$\phi_{\rm H} =$$-129^{+30}_{-29}$,  	$\phi_{\rm C} =$$-111^{+32}_{-33}$,  	$\phi_{\rm HI} =$$-27879^{+65}_{-65}$,  	$\sigma_{\rm S} =$$9^{+11}_{-6.2}$,  	$\sigma_{\rm S} =$$9^{+11}_{-5.9}$,  	$\sigma_{\rm H} =$$3.9^{+2.7}_{-2.2}$,  	$\sigma_{\rm C} =$$4.6^{+5.7}_{-3.2}$,  	$\sigma_{\rm HI} =$$9^{+12}_{-6.2}$ 	 \\ 
wasp-35b             & 	$0.28^{+0.36}_{-0.20}$ & 	$57^{+32}_{-35}$ & 	$48^{+34}_{-31}$ & 	$-0.1^{+3.1}_{-2.9}$ & 	$-2^{+26}_{-24}$ & 	$\phi_{\rm FI} =$$19^{+36}_{-32}$,  	$\sigma_{\rm C} =$$16^{+14}_{-8.9}$,  	$\sigma_{\rm FI} =$$25^{+37}_{-18}$ 	 \\ 
wasp-36b             & 	$0.20^{+0.16}_{-0.10}$ & 	$53^{+31}_{-31}$ & 	$57^{+31}_{-34}$ & 	$-0.7^{+2.9}_{-2.4}$ & 	$0^{+21}_{-21}$ & 	$\phi_{\rm H} =$$13^{+38}_{-39}$,  	$\sigma_{\rm C} =$$7.8^{+8.6}_{-5.5}$,  	$\sigma_{\rm H} =$$24.9^{+7.6}_{-5.1}$ 	 \\ 
wasp-42b             & 	$0.136^{+0.057}_{-0.032}$ & 	$69^{+22}_{-29}$ & 	$56^{+30}_{-31}$ & 	$-1.9^{+1.7}_{-1.6}$ & 	$-5^{+14}_{-14}$ & 	$\phi_{\rm H} =$$-24^{+13}_{-13}$,  	$\sigma_{\rm C} =$$3.2^{+3.5}_{-2.3}$,  	$\sigma_{\rm H} =$$1.5^{+1.6}_{-1.0}$ 	 \\ 
wasp-43b             & 	$0.52^{+0.46}_{-0.40}$ & 	$41^{+38}_{-28}$ & 	$58^{+30}_{-34}$ & 	$1.1^{+1.7}_{-3.2}$ & 	$7^{+13}_{-13}$ & 	$\phi_{\rm H} =$$11^{+45}_{-45}$,  	$\sigma_{\rm C} =$$10^{+13}_{-7.1}$,  	$\sigma_{\rm H} =$$3.0^{+3.5}_{-2.1}$ 	 \\ 
wasp-46b             & 	$0.44^{+0.48}_{-0.44}$ & 	$49^{+35}_{-30}$ & 	$51^{+34}_{-32}$ & 	$0.1^{+2.6}_{-2.9}$ & 	$1^{+25}_{-25}$ & 	$\sigma_{\rm C} =$$18^{+17}_{-12}$ 	 \\ 
wasp-49b             & 	$0.163^{+0.095}_{-0.056}$ & 	$52^{+33}_{-34}$ & 	$50^{+34}_{-32}$ & 	$0.8^{+2.1}_{-3.4}$ & 	$-6^{+10}_{-11}$ & 	$\sigma_{\rm C} =$$6.9^{+5.1}_{-4.7}$ 	 \\ 
wasp-4b              & 	$0.22^{+0.15}_{-0.12}$ & 	$57^{+28}_{-31}$ & 	$58^{+28}_{-32}$ & 	$-1.6^{+2.2}_{-1.9}$ & 	$-1.9^{+9.6}_{-8.1}$ & 	$\phi_{\rm C} =$$-10.4^{+7.4}_{-7.2}$,  	$\phi_{\rm H} =$$-60^{+18}_{-16}$,  	$\phi_{\rm HI} =$$57799^{+45}_{-37}$,  	$\sigma_{\rm C} =$$7.5^{+7.8}_{-5.1}$,  	$\sigma_{\rm C} =$$3.9^{+4.9}_{-2.8}$,  	$\sigma_{\rm H} =$$7.8^{+2.6}_{-2.3}$,  	$\sigma_{\rm HI} =$$9^{+20}_{-6.7}$ 	 \\ 
wasp-50b             & 	$0.28^{+0.40}_{-0.20}$ & 	$54^{+31}_{-32}$ & 	$57^{+30}_{-35}$ & 	$-0.7^{+3.1}_{-2.4}$ & 	$-2^{+26}_{-23}$ & 	$\sigma_{\rm C} =$$10.7^{+6.9}_{-6.1}$ 	 \\ 
wasp-52b             & 	$0.142^{+0.076}_{-0.037}$ & 	$80.6^{+0.95}_{-2.0}$ & 	$8.2^{+8.2}_{-2.6}$ & 	$-2.3^{+1.9}_{-0.93}$ & 	$23^{+9.6}_{-15}$ & 	$\phi_{\rm S} =$$-13^{+20}_{-22}$,  	$\phi_{\rm S} =$$142^{+12}_{-12}$,  	$\phi_{\rm S} =$$41^{+11}_{-11}$,  	$\phi_{\rm S} =$$8^{+14}_{-14}$,  	$\sigma_{\rm C} =$$5.1^{+6.2}_{-3.6}$,  	$\sigma_{\rm S} =$$14^{+21}_{-10}$,  	$\sigma_{\rm S} =$$6.4^{+5.7}_{-4.4}$,  	$\sigma_{\rm S} =$$8^{+11}_{-5.9}$,  	$\sigma_{\rm S} =$$8^{+11}_{-6.0}$ 	 \\ 
wasp-5b              & 	$0.18^{+0.12}_{-0.076}$ & 	$30^{+47}_{-19}$ & 	$58^{+29}_{-34}$ & 	$0.6^{+1.1}_{-2.7}$ & 	$-22^{+15}_{-9.3}$ & 	$\phi_{\rm H} =$$-27^{+14}_{-13}$,  	$\sigma_{\rm C} =$$6.0^{+6.1}_{-4.2}$,  	$\sigma_{\rm H} =$$1.7^{+1.5}_{-1.1}$ 	 \\ 
wasp-65b             & 	$0.17^{+0.11}_{-0.070}$ & 	$49^{+33}_{-30}$ & 	$55^{+31}_{-33}$ & 	$-0.0^{+2.9}_{-2.8}$ & 	$-3^{+18}_{-17}$ & 	$\sigma_{\rm C} =$$5.6^{+6.2}_{-3.9}$ 	 \\ 
wasp-68b             & 	$0.110^{+0.017}_{-0.0079}$ & 	$51^{+34}_{-33}$ & 	$45^{+36}_{-30}$ & 	$2.0^{+1.5}_{-3.5}$ & 	$13.9^{+3.0}_{-3.3}$ & 	$\sigma_{\rm C} =$$3.2^{+2.9}_{-2.2}$ 	 \\ 
wasp-77ab            & 	$0.24^{+0.17}_{-0.14}$ & 	$52^{+32}_{-31}$ & 	$54^{+32}_{-31}$ & 	$-1.0^{+2.5}_{-2.2}$ & 	$0.7^{+8.2}_{-8.4}$ & 	$\phi_{\rm H} =$$-51^{+16}_{-17}$,  	$\sigma_{\rm C} =$$9.2^{+8.6}_{-5.9}$,  	$\sigma_{\rm H} =$$3.4^{+4.5}_{-2.4}$ 	 \\ 
wasp-80b             & 	$0.130^{+0.048}_{-0.025}$ & 	$53^{+33}_{-33}$ & 	$52^{+32}_{-33}$ & 	$-0.8^{+2.8}_{-2.4}$ & 	$-3^{+14}_{-15}$ & 	$\phi_{\rm H} =$$1.2^{+9.4}_{-9.2}$,  	$\sigma_{\rm C} =$$3.7^{+4.1}_{-2.6}$,  	$\sigma_{\rm H} =$$4.2^{+5.3}_{-2.9}$ 	 \\ 
wasp-98b             & 	$0.29^{+0.30}_{-0.21}$ & 	$53^{+32}_{-32}$ & 	$55^{+31}_{-34}$ & 	$0.1^{+2.8}_{-3.0}$ & 	$3^{+22}_{-25}$ & 	$\sigma_{\rm C} =$$14^{+15}_{-10}$ 	 \\ 
wts-2b               & 	$1.8^{+4.0}_{-3.7}$ & 	$50^{+33}_{-31}$ & 	$49^{+34}_{-30}$ & 	$-0.4^{+3.0}_{-2.6}$ & 	$4^{+22}_{-27}$ & 	$\sigma_{\rm HRS} =$$41^{+33}_{-29}$ 	 \\

\hline
\hline
\end{tabular}
\tablefoot{\scriptsize Instrumental offsets refer to the first instrument used and instrument codes are as follows: H=HARPS, HN=HARPS-N, HI=HIRES, C=CORALIE, PFS=PFS, T=TRES, S=SOPHIE, F=FEROS, SU=Subaru, CA=CAFE, CY=CYCLOPS, U=UCLES, FL=FLAMES, FI=FIES, HRS=HRS}
\end{center}
\label{tab:results}
\end{table}%
\end{landscape}

\begin{table*}[]
\setlength{\extrarowheight}{4pt}
\caption{Maximum mass of possible trojan bodies for the six tested models assuming their presence. We present the 99.7\% confidence intervals of the mass computed from random samplings of the radial velocity semi-amplitude $K_2$, the inclination $i$, the eccentricity $e$ (when applicable), and the stellar mass obtained from the literature. The values listed in this table are shown in Fig.\ref{fig:maxmass}}
\scriptsize
\begin{center}
\begin{tabular}{l cc|cc|c}
\hline
\hline
       &        & $m_t^{\dagger}$  & \multicolumn{2}{|c|}{2$-\sigma$ upper mass limit ($M_{\oplus}$)} &  \\ \hline

System    & $\alpha$  &  ($M_{\oplus}$)  & L$_4$ & L$_5$  & $\alpha/\sigma_{\alpha}$    \\
\hline

corot-1b             & 	$-0.028^{+0.074}_{-0.077}$ 	 & $-11^{+28}_{-29}$ 	 & $<-71.0$ & $<50.2$ & $0.4$ \\ 
corot-2b             & 	$-0.021^{+0.059}_{-0.060}$ 	 & $-26^{+72}_{-73}$ 	 & $<-167.4$ & $<126.3$ & $0.4$ \\ 
corot-15b            & 	$0.011^{+0.034}_{-0.032}$ 	 & $250^{+780}_{-760}$ 	 & $<-1352.3$ & $<1763.4$ & $0.3$ \\ 
corot-18b            & 	$0.039^{+0.046}_{-0.045}$ 	 & $49^{+58}_{-58}$ 	 & $<-82.0$ & $<167.3$ & $0.9$ \\ 
epic211089792b       & 	$-0.15^{+0.10}_{-0.11}$ 	 & $-39^{+27}_{-31}$ 	 & $<-93.5$ & $<19.5$ & $1.4$ \\ 
gj3470b              & 	$-0.61^{+0.23}_{-0.26}$ 	 & $-9.8^{+3.6}_{-4.2}$ 	 & $<-15.6$ & $<-1.3$ & $2.5$ \\ 
hat-p-12b            & 	$-0.18^{+0.14}_{-0.12}$ 	 & $-14^{+11}_{-8.9}$ 	 & $<-39.4$ & $<7.2$ & $1.4$ \\ 
hat-p-20b            & 	$0.000^{+0.013}_{-0.013}$ 	 & $1^{+35}_{-35}$ 	 & $<-72.4$ & $<72.3$ & $0.0$ \\ 
hat-p-23b            & 	$-0.120^{+0.096}_{-0.095}$ 	 & $-92^{+74}_{-73}$ 	 & $<-237.7$ & $<55.6$ & $1.3$ \\ 
hat-p-36b            & 	$0.25^{+0.22}_{-0.24}$ 	 & $170^{+150}_{-160}$ 	 & $<-150.5$ & $<530.4$ & $1.1$ \\ 
hat-p-54b            & 	$0.14^{+0.72}_{-0.59}$ 	 & $40^{+200}_{-160}$ 	 & $<-258.0$ & $<268.2$ & $0.2$ \\ 
hats-2b              & 	$0.05^{+0.50}_{-0.46}$ 	 & $20^{+250}_{-230}$ 	 & $<-427.7$ & $<416.8$ & $0.1$ \\ 
hats-9b              & 	$-0.26^{+0.42}_{-0.46}$ 	 & $-80^{+130}_{-140}$ 	 & $<-287.4$ & $<221.6$ & $0.6$ \\ 
k2-30b               & 	$0.02^{+0.28}_{-0.25}$ 	 & $4^{+64}_{-57}$ 	 & $<-163.8$ & $<141.5$ & $0.1$ \\ 
kepler-17b           & 	$0.033^{+0.072}_{-0.074}$ 	 & $29^{+65}_{-66}$ 	 & $<-103.5$ & $<160.2$ & $0.4$ \\ 
ogle-tr-113b         & 	$-0.13^{+0.29}_{-0.26}$ 	 & $-57.64^{+130}_{-120}$ 	 & $<-362.7$ & $<238.4$ & $0.5$ \\ 
ogle-tr-56b          & 	$-0.19^{+0.26}_{-0.22}$ 	 & $-90^{+120}_{-110}$ 	 & $<-376.3$ & $<130.1$ & $0.8$ \\ 
qatar-1b             & 	$-0.008^{+0.077}_{-0.095}$ 	 & $-3^{+31}_{-38}$ 	 & $<-67.6$ & $<79.0$ & $0.1$ \\ 
qatar-2b             & 	$-0.028^{+0.038}_{-0.038}$ 	 & $-25^{+35}_{-34}$ 	 & $<-92.5$ & $<46.6$ & $0.7$ \\ 
tres-3b              & 	$-0.066^{+0.035}_{-0.036}$ 	 & $-46^{+25}_{-25}$ 	 & $<-105.6$ & $<14.1$ & $1.8$ \\ 
tres-5b              & 	$-0.03^{+0.53}_{-0.53}$ 	 & $-20^{+340}_{-350}$ 	 & $<-603.0$ & $<576.2$ & $0.1$ \\ 
wasp-103b            & 	$-0.12^{+0.29}_{-0.29}$ 	 & $-60^{+160}_{-160}$ 	 & $<-399.1$ & $<277.3$ & $0.4$ \\ 
wasp-104b            & 	$0.004^{+0.070}_{-0.071}$ 	 & $2^{+33}_{-33}$ 	 & $<-67.8$ & $<76.7$ & $0.1$ \\ 
wasp-12b             & 	$0.015^{+0.028}_{-0.027}$ 	 & $8^{+15}_{-14}$ 	 & $<-23.2$ & $<33.6$ & $0.5$ \\ 
wasp-135b            & 	$-0.002^{+0.060}_{-0.060}$ 	 & $-1^{+42}_{-42}$ 	 & $<-87.7$ & $<87.8$ & $0.0$ \\ 
wasp-13b             & 	$-0.24^{+0.49}_{-0.66}$ 	 & $-44^{+87}_{-120}$ 	 & $<-170.2$ & $<158.5$ & $0.4$ \\ 
wasp-18b             & 	$-0.0022^{+0.0090}_{-0.0083}$ 	 & $-9^{+35}_{-32}$ 	 & $<-79.6$ & $<59.7$ & $0.3$ \\ 
wasp-19b             & 	$0.034^{+0.034}_{-0.033}$ 	 & $14^{+14}_{-14}$ 	 & $<-14.2$ & $<43.1$ & $1.0$ \\ 
wasp-23b             & 	$-0.001^{+0.052}_{-0.054}$ 	 & $-0^{+17}_{-18}$ 	 & $<-33.6$ & $<34.6$ & $0.0$ \\ 
wasp-2b              & 	$-0.011^{+0.019}_{-0.025}$ 	 & $-3.6^{+5.8}_{-7.8}$ 	 & $<-15.3$ & $<14.6$ & $0.5$ \\ 
wasp-35b             & 	$0.08^{+0.20}_{-0.21}$ 	 & $21^{+54}_{-55}$ 	 & $<-136.6$ & $<159.9$ & $0.4$ \\ 
wasp-36b             & 	$0.092^{+0.043}_{-0.043}$ 	 & $77^{+36}_{-36}$ 	 & $<5.4$ & $<145.8$ & $2.2$ \\ 
wasp-42b             & 	$0.015^{+0.058}_{-0.057}$ 	 & $3^{+11}_{-10}$ 	 & $<-18.4$ & $<23.7$ & $0.3$ \\ 
wasp-43b             & 	$-0.000^{+0.029}_{-0.027}$ 	 & $-0^{+22}_{-20}$ 	 & $<-43.3$ & $<39.9$ & $0.0$ \\ 
wasp-46b             & 	$-0.012^{+0.055}_{-0.053}$ 	 & $-9^{+43}_{-41}$ 	 & $<-101.2$ & $<77.9$ & $0.2$ \\ 
wasp-49b             & 	$-0.03^{+0.18}_{-0.17}$ 	 & $-4^{+25}_{-24}$ 	 & $<-62.7$ & $<49.9$ & $0.2$ \\ 
wasp-4b              & 	$-0.021^{+0.020}_{-0.019}$ 	 & $-9.7^{+9.3}_{-8.6}$ 	 & $<-28.7$ & $<7.2$ & $1.1$ \\ 
wasp-50b             & 	$0.02^{+0.10}_{-0.096}$ 	 & $9^{+53}_{-51}$ 	 & $<-131.8$ & $<128.1$ & $0.2$ \\ 
wasp-52b             & 	$0.01^{+0.11}_{-0.11}$ 	 & $1^{+18}_{-18}$ 	 & $<-34.5$ & $<41.0$ & $0.1$ \\ 
wasp-5b              & 	$0.005^{+0.013}_{-0.012}$ 	 & $3.0^{+7.7}_{-7.1}$ 	 & $<-13.4$ & $<17.9$ & $0.4$ \\ 
wasp-65b             & 	$-0.012^{+0.075}_{-0.077}$ 	 & $-7^{+43}_{-44}$ 	 & $<-93.3$ & $<84.2$ & $0.2$ \\ 
wasp-68b             & 	$0.040^{+0.060}_{-0.067}$ 	 & $14^{+21}_{-23}$ 	 & $<-30.1$ & $<60.3$ & $0.6$ \\ 
wasp-77ab            & 	$0.045^{+0.037}_{-0.040}$ 	 & $29^{+24}_{-26}$ 	 & $<-24.4$ & $<92.9$ & $1.2$ \\ 
wasp-80b             & 	$0.038^{+0.083}_{-0.085}$ 	 & $8^{+17}_{-17}$ 	 & $<-28.5$ & $<44.7$ & $0.5$ \\ 
wasp-98b             & 	$0.14^{+0.48}_{-0.46}$ 	 & $40^{+150}_{-140}$ 	 & $<-241.2$ & $<279.0$ & $0.3$ \\ 
wts-2b               & 	$0.22^{+0.72}_{-0.56}$ 	 & $90^{+300}_{-230}$ 	 & $<-371.2$ & $<399.4$ & $0.3$ \\

\hline
\hline
\end{tabular}
\tablefoot{$\dagger$ Estimation of the trojan mass based on $\alpha$ and assuming small eccentricity, $m_t<<m_p$, and trojan location around one the two stable Lagrangian points. Negative values correspond to $L_4$ and positive values correspond to $L_5$}
\end{center}
\label{tab:maxmass}
\end{table*}%

\end{document}